\documentclass[12pt]{article}
\usepackage{epsfig,amsmath}

\begin{document}

\title{Rotating relativistic thin disks as sources of charged
and magnetized Kerr-NUT spacetimes}

\author{Gonzalo Garc\'{\i}a-Reyes
\thanks{e-mail: ggarcia@utp.edu.co}   \\
{\it Universidad  Tecnol\'ogica de Pereira, Departamento de 
F\'{\i}sica, }   \\
{\it A. A. 097, Pereira, Colombia }  \\
\and
Guillermo A. Gonz\'{a}lez
\thanks{e-mail: guillego@uis.edu.co} \\
{\it Escuela de F\'{\i}sica, Universidad Industrial de Santander,}	\\
{\it A.A. 678, Bucaramanga, Colombia }  }

\date{ }

\maketitle

\begin{abstract}

A family  of models of  counterrotating and rotating   relativistic thin disks  of infinite extension   based on a charged and magnetized Kerr-NUT metric are constructed 
using the well-known ``displace, cut and
reflect'' method extended to solutions of vacuum Einstein-Maxwell equations. The metric  considered has as
limiting cases a charged and magnetized Taub-NUT solution 
and the well known  Kerr-Newman solutions. We show that for 
Kerr-Newman fields the eigenvalues of the
energy-momentum  tensor of the disk are for all the
values of the parameters real quantities 
so that these  disks do not present  heat flow in any case, whereas for charged and magnetized 
Kerr-NUT and Taub-NUT fields we find always
regions  with  heat flow. We also find a general constraint over the counterrotating tangential velocities needed to cast the surface  energy-momentum tensor of the disk as the superposition of
two counterrotating charged dust fluids. 
We show that, in general, it is
not possible to take the two counterrotating fluids as circulating along
electrogeodesics nor take the two counterrotating tangential 
velocities as equal and opposite.


\end{abstract}

\newpage


\section{Introduction}
	
Stationary or static axially symmetric  exact solutions to the Einstein  field equations
representing  rela\-tivistic thin disks are of  great astrophysical importance since
they can be used  as  models for certain galaxies, accretion disks, and the superposition of a black holes and a galaxy or an accretion disk as in the case of quasars.
Disk sources for stationary axially symmetric spacetimes  with  electromagnetic fields, especially magnetic fields, 
are also of astrophysical importance in the study of neutron stars, white dwarfs and  galaxy formation. 
In such situation one has to study the coupled Einstein-Maxwell equations.

Exact solutions which describe relativistic static thin disks without radial pressure
were first studied by Bonnor and Sackfield \cite{BS}, 
and Morgan and Morgan \cite{MM1}, and with radial  pressure by Morgan and Morgan \cite{MM2}.  Also thin disks with radial tension were considered \cite{GL1}.  
Several classes of exact solutions of the  Einstein field  equations
corresponding to static thin disks with or  without radial pressure have been
obtained by different authors \cite{LP}--\cite{GE}. 
Rotating  thin disks that can be considered as a source of a Kerr metric were
presented by  Bi\u{c}\'ak and  Ledvinka \cite{BL}, while rotating disks with
heat flow were were studied by Gonz\'alez and Letelier \cite{GL2}. The
nonlinear  superposition of a disk and a black hole was first obtained by Lemos
and Letelier \cite{LL1}. Perfect fluid disks with halos were studied by Vogt				
and Letelier \cite{VL1}. The stability of some general relativistic thin disks
models using a first order perturbation of the energy-momentum tensor was
investigated by Ujevic and Letelier \cite{UL1}.  

Thin disks in precense of electromagnetic field  have been discussed as sources for
Kerr-Newman fields \cite{LBZ,GG4}, magnetostatic  axisymmetric fields \cite{LET1},
conformastationary metrics \cite{KBL},  while models of electrovacuum static
counterrotating dust disks  were presented in \cite{GG1}. Charged perfect fluid
disks were also studied by Vogt and Letelier \cite{VL2}, and  charged perfect
fluid disks as sources of  static and  Taub-NUT-type spacetimes by Garc\'\i
a-Reyes and Gonz\'alez \cite{GG2,GG3}.  

In all the above cases, the disks are obtained by an ``inverse problem''
approach, called by Synge the ``{\it g-method}'' \cite{SYN}. The method works
as follows: a solution of the vacuum Einstein equations is taken, such that
there is a discontinuity in the derivatives of the metric tensor on the plane
of the disk,  and the energy-momentum tensor is obtained from the Einstein
equations. The physical properties of the matter  distribution are then studied
by an analysis of the surface energy-momentum tensor so obtained. Another
approach to  generate disks is by solving the Einstein equations given a source
(energy-momentum tensor). Essentially, they are obtained by solving a
Riemann-Hilbert problem and are highly nontrivial 
\cite{NM}--\cite{KLE4}.
A review of this kind of disks solutions to the Einstein-Maxwell
equations was presented by Klein in \cite{KLE5}.

Now, when the inverse problem approach is used for static spacetimes, the energy-momentum tensor is diagonal and its analysis is direct
and, except for  dust disks,  the solutions obtained  have anisotropic sources
with azimuthal stress  different from the radial stress. On the other hand,
when the considered spacetime is stationary, the obtained energy-momentum
tensor is non-diagonal and the  analysis of its physical content is more
involved and, in general, the obtained source is not only anisotropic but with
nonzero heat flow. Due to this fact, there are  very few  works about of stationary  disks and they are limited to disks without heat flow 
\cite{BL,LBZ,GG3}. Indeed, only in one work disks with
heat flow has been considered, but only a partial analysis
of the corresponding energy-momentum tensor was made
\cite{GL2}.  

The above disks can also be interpreted as made of two counterrotating streams of freely moving (charged) particles, i.e., which move along (electro-)geodesics, as was also indicated in \cite{LBZ}. This interpretation is obtained by means of the Counterrotating Model (CRM) in which the energy-momentum tensor of the source is expressed as the
superposition of two counterrotating  fluids. Now, even though this
interpretation can be seen as merely theoretical, there are observational
evidence of disks made of  counterrotating matter as in the
case of certain $S0$ and  $Sa$ galaxies
\cite{RGK}--\cite{STRUCK}. Has also been observed in some spiral galaxies as in NGC3626 the presence of counterrotating ionized (charged) gas disks \cite{CBG}. Indeed recent investigations have shown that there is a large number of galaxies (the first was NGC4550 in Virgo) \cite{STRUCK}  which show counterrotating streams  in the disk with up to $50\%$ counterrotation.
It is believe that the presence of counterrotating matter components in these galaxies is the consecuence of the accretion or merger of galaxies. Thus
these counterrotating streams may be the result from the capture by a massive early-type galaxy of a gas-rich dwarf galaxy that was orbiting in the opposite sense to the rotation of the main galaxy.

The purpose of the present paper is twofold. In first instance, we present a
analysis of the energy-momentum tensor and the surface current density
for electrovacuum stationary axially symmetric relativistic thin disks of infinite extension  without
radial stress and when there is  heat flow. And, in the second place, we
present a study of the counterrotating model (CRM) for these stationary
thin disks. The paper is  structured as follows. 
In Sec. II  we present a
summary of the procedure to obtain  models of  rotating thin disks  with a
purely azimuthal pressure  and  currents, using the well
known ``displace, cut 
and reflect'' method extended to solutions of  Einstein-Maxwell equations. In particular, we obtain expressions for the surface 
energy-momentum tensor  and the surface current density of the disks.   

In Sec. III the disks are interpreted in terms of  the CRM. 
We  find a general constraint over the counterrotating tangential velocities needed to cast the surface  energy-momentum tensor of the disk as the superposition of
two counterrotating charged  fluids made of dust or pressureless matter (collisionless particles). 
We show that, in general, it is
not possible to take the two counterrotating fluids as circulating along
electro-geodesics nor take the two counterrotating tangential 
velocities as equal and opposite. 

In the following section, Sec. IV, a family
of models of counterrotaing  and  rotating relativistic thin disks based on a 
magnetized and charged Kerr-NUT metric is considered.
This metric has as
limiting cases a charged and magnetized Taub-NUT solution 
and the well known  Kerr-Newman solutions. We show that for 
Kerr-Newman fields the eigenvalues of the
energy-momentum  tensor of the disk are for all the
values of the parameters real quantities 
so that this disks do not present  heat flow in any case, whereas for
charged and magnetized 
Kerr-NUT and Taub-NUT fields we find always
regions  with  heat flow. 
We then analyze the CRM 
for these disks and we study the  angular velocities,
surface energy densities  and electric charge densities  of both  streams when 
the two fluids move
along electrogeodesics and when they move with equal 
and opposite  velocities.
Also the stability against radial  perturbation is 
analyzed in  all the
cases. Finally, in Sec. V, we summarize our main results.   

  
\section{Electrovacuum rotating relativistic thin disks}

A sufficiently general metric for our purposes can be  written as the
Weyl-Lewis-Papapetrou line element \cite{KSHM},   
\begin{equation}
ds^2 = - \ e^{2 \Psi} (dt + {\cal W} d\varphi)^2 \ +  \ e^{- 2 \Psi}[ r^2 
d\varphi^2 + e^{2 \Lambda} (dr^2 + dz^2)], \label{eq:met}
\end{equation}
where we use for the coordinates the notation $(x^0,x^1,x^2,x^3) =
(t,\varphi,r,z)$, and $\Psi$, ${\cal W}$, and $\Lambda$ are functions of $r$
and $z$ only.  The vacuum Einstein-Maxwell  equations,  in  geometric units in
which $8 \pi G = c = \mu _0 =  \varepsilon _0 = 1$,  are given by 
\begin{subequations}\begin{eqnarray}
&   &   G_{ab} \  =  \ T_{ab},  \label{eq:einmax1}    \\
&   &    F^{ab}_{ \ \ \ ; b} \ =  \ 0 \label{eq:einmax2},     
\end{eqnarray}\end{subequations}
with
\begin{subequations}\begin{eqnarray}
T_{ab} \  &=&  \ F_{ac}F_b^{ \ c} - \frac{1}{4} g_{ab}  
F_{cd} F^{cd}, \label{eq:et}  \\
F_{ab} \ &=&  \  A_{b,a} -  A_{a,b},
\end{eqnarray}\end{subequations} 
where $A_a = (A_t, A_\varphi, 0, 0)$ and the electromagnetic  potentials  $A_t$
and $A_\varphi$ are also functions of  $r$ and $z$  only.

For the metric (\ref{eq:met}), the Einstein-Maxwell equations are equivalent to
the system \cite{E2}
\begin{subequations}\begin{eqnarray}
&& \qquad \qquad \nabla \cdot [ r^{-2}f( \nabla A_\varphi - {\cal W} \ \nabla
A_t ]=0, \label{em11} \\
&&  \qquad \quad \nabla \cdot [ f^{-1} \nabla A_t + r^{-2} f {\cal W} ( \nabla
A_\varphi - {\cal W} \ \nabla A_t ]=0, \label{em12} \\
&& \qquad \nabla \cdot [r^{-2} f^{2} \nabla {\cal W} - 2 r^{-2}
f A_t (\nabla A_\varphi - {\cal W} \ \nabla A_t)] = 0 , \label{em13} \\
&&f \nabla^{2} f  = \nabla f \cdot \nabla f - r^{-2} f^{4} \nabla{\cal W}
\cdot  \nabla{\cal W} + f \nabla A_t \cdot \nabla A_t  \nonumber \\
&& \qquad \qquad + \ r^{-2} f^{3} (\nabla A_\varphi - {\cal W} \nabla A_t)
\cdot (\nabla A_\varphi - {\cal W} \nabla A_t) , \label{em14} \\
\Lambda_{,r} &=& r (\Psi_{,r}^2 - \Psi_{,z}^2 )
- \frac 1 {4r} ({\cal W}_{,r}^2 -{\cal W}_{,z}^2)e^{4 \Psi }
- \frac 1 {2r} (r^2 e^{-2 \Psi }-{\cal W}^2e^{2 \Psi }) 
(A_{t,r}^2-A_{t,z}^2) \nonumber  \\
&&+ \ \frac 1 {2r}  (A_{\varphi,r}^2-A_{\varphi,z}^2) e^{2 \Psi }
-\frac{1}{r}{\cal W}(A_{\varphi,r} A_{t,r}-
A_{\varphi,z} A_{t,z})e^{2 \Psi}, \\
\Lambda_{,z} &=&  2r \Psi_{,r} \Psi_{,z} - \frac{1}{2r} 
{\cal W}_{,r} {\cal W}_{,z} e^{4 \Psi} -\frac{1}{r}(r^2
e^{-2 \Psi} -{\cal W}^2 e^{2 \Psi})A_{t,r} A_{t,z} \nonumber \\
&& + \ \frac{1}{r}A_{\varphi,r} A_{\varphi,z} e^{2 \Psi} 
 -\frac{1}{r}{\cal W}(A_{\varphi,r} A_{t,z}+
A_{\varphi,z} A_{t,r})e^{2 \Psi},  \label{eq:lamz}
\end{eqnarray}\end{subequations}
where $\nabla$ is the  standard  differential 
operator in cylindrical coordinates and $f = e^{2 \Psi}$. 

In order to obtain a solution of (\ref{eq:einmax1}) - (\ref {eq:einmax2}) 
representing a thin disk at $z=0$, we assume  that the components of the metric
tensor are continuous across the disk, but with first  derivatives
discontinuous on the  plane $z=0$, with  discontinuity functions 
\begin{eqnarray}
b_{ab} \ &=& g_{ab,z}|_{_{z = 0^+}} \ - \ g_{ab,z}|_{_{z = 0^-}} \ = \ 2 \
 g_{ab,z}|_{_{z = 0^+}}.                  
\end{eqnarray}
Thus, by using the distributional approach \cite{PH,LICH,TAUB} or the junction
conditions on the extrinsic curvature of thin shells \cite{IS1,IS2,POI}, the
Einstein-Maxwell equations yield an  energy-momentum tensor $T_{ab}=
T^{\mathrm{elm}}_{ab} +  T^{\mathrm {mat}}_{ ab}$,  where 
$T^{\mathrm{mat}}_{ab} = Q_{ab} \ \delta (z)$, and a  current density $J_a =
j_a  \delta (z)= - 2 e^{2 (\Psi - \Lambda)} A_{a,z} \delta (z)$, where $\delta
(z)$ is the  usual Dirac function with support on  the disk.  $T^{\mathrm {
elm}}_{ab}$ is the electromagnetic tensor defined in Eq. (\ref{eq:et}), $j_a$
is the current density on the plane  $z=0$, and 
$$
Q^a_b = \frac{1}{2}\{b^{az}\delta^z_b - b^{zz}\delta^a_b +  g^{az}b^z_b -
g^{zz}b^a_b + b^c_c (g^{zz}\delta^a_b - g^{az}\delta^z_b)\}
$$
is the distributional energy-momentum tensor. The ``true''  surface
energy-momentum tensor (SEMT) of the  disk,  $S_{ab}$, and the ``true'' surface
current density,  $\mbox{\sl j}_a$, can be obtained through the
relations  	
\begin{subequations}\begin{eqnarray}
S_{ab} &=& \int T^{\mathrm {mat}}_{ab} \ ds_n \ = \ e^{  \Lambda - \Psi} \ 
Q_{ab} \ ,   \\
\mbox{\sl j}_a  &=& \int J_{a}  \ ds_n \ = \ e^{ \Lambda -  \Psi} \ j_a \ , 
\end{eqnarray}\end{subequations}
where $ds_n = \sqrt{g_{zz}} \ dz$ is the ``physical  measure'' of length in the
direction normal to the disk.

For the metric (\ref{eq:met}), the nonzero components of  $S_a^b$ are
\begin{subequations}\begin{eqnarray}
&S^0_0 &= \ \frac{e^{\Psi - \Lambda}}{ r^2} \left [ 2 r^2 (\Lambda,_z - \ 2
\Psi,_z) - \ e^{4\Psi} {\cal W} {\cal W},_z \right ] , \label{eq:emt1}  \\
&S^0_1 &= \ - \frac{e^{\Psi - \Lambda}}{ r^2} \left [  4 r^2 {\cal W}  \Psi,_z
+ \ ( r^2 + \ {\cal W}^2 e^{4\Psi} ) {\cal W},_z \right ] , \label{eq:emt2}  \\
&S^1_0 &= \ \frac{e^{\Psi - \Lambda}}{ r^2}  \left[ e^{4\Psi}  {\cal W},_z
\right ], \label{eq:emt3} \\
&S^1_1 &= \ \frac{e^{\Psi - \Lambda}}{ r^2} \left [ 2 r^2 \Lambda,_z + \
e^{4\Psi} {\cal W} {\cal W},_z \right ] ,  \label{eq:emt4} 
\end{eqnarray}\end{subequations} 
and the nonzero components of the surface current density  $\mbox{\sl j}_a$
are   
\begin{subequations}\begin{eqnarray}
& \mbox{\sl j}_t &= \ -2 e^{\Psi - \Lambda} A _{t,z} , 
\label{eq:corelec}   \\
&\mbox{\sl j}_{\varphi} &= \ -2 e^{\Psi - \Lambda} 
A _{\varphi ,z},  \label{eq:cormag} 
\end{eqnarray}\end{subequations}
where all the quantities are evaluated at $z = 0^+$.

These disks are essentially of  infinite extension. Finite disks can be
obtained  introducing oblate  spheroidal coordinates,  which are  naturally
adapted to a disk source, and  imposing  appropriate boundary conditions. These
solutions, in the vacuum and static case, correspond to the Morgan and Morgan
solutions \cite{MM1}. A more general class of  solutions representating finite
thin disks can be constructed  using a method based on the use of conformal
transformations and solving a boundary-value problem  
\cite{MM2,CHGS,GL1,GE,GG2,GG3}. 
 
Now, in order to analyze the matter content of the disks is   necessary to
compute the eigenvalues and eigenvectors of  the energy-momentum tensor. The
eigenvalue problem for the SEMT (\ref{eq:emt1}) - (\ref{eq:emt4})
\begin{equation}
S^a_b \ \xi^b \ = \lambda \ \xi^a,
\end{equation}
has the solutions
\begin{equation}
\lambda_\pm \ = \ \frac{1}{2} \left( \ T \pm \sqrt{D} \  \right) , 
\end{equation}
where
\begin{equation}
T = S^0_0 \ + \ S^1_1 \ , \quad D = ( S^1_1 - S^0_0 )^2 +  4 \ S^0_1 \ S^1_0 ,
\end{equation}
and $\lambda_r = \lambda_z = 0$. For the metric  (\ref{eq:met})    
\begin{eqnarray}
D  & = & 4 \frac{e^{2(\Psi - \Lambda)}}{r^2} (4r^2  \Psi_{,z}^2-{\cal W}_{,z}^2
e^{4\Psi} ) = A^2 - B^2, \\
T &=& 4 e^{\Psi - \Lambda}(\Lambda_{,z} - \Psi_{,z}),
\end{eqnarray}  
where 
\begin{equation}
A=4 \Psi_{,z}e^{\Psi - \Lambda} , \ \ \ \ B = \frac 2 r   {\cal W}_{,z}
e^{3\Psi -\Lambda}.
\end{equation}
The corresponding eigenvectors are
\begin{equation}\begin{array}{ccl}
\xi^a_\pm & = &  (\ \xi^0_\pm  , \ \xi^1_\pm  , \ 0, \ 0 ), \\
	&	&	\\
X^a & = & e^{\Psi - \Lambda} ( 0, 0, 1, 0 ),	\\
	&	&	\\
Y^a & = & e^{\Psi - \Lambda} ( 0, 0, 0, 1 ),
\end{array}\label{eq:tetrad}\end{equation}
with
\begin{equation}
g (\xi_\pm, \xi_\pm ) = 2  N_\pm e^{2\Psi} \left (\frac  {\xi^0_\pm }{S^0_0 -
S^1_1 \pm \sqrt D } \right )^2    , 
\end{equation}
where
\begin{equation}
 N_\pm  = \sqrt { D} (-\sqrt { D} \pm A). \label{eq:norm}
\end{equation}

We only consider the case when $D \geq 0$, so that the two eigenvalues
$\lambda_\pm$ are real  and different and the two eigenvectors are orthogonal, 
in such a way that one of them is timelike and the other  is spacelike. Since
$|A|\geq \sqrt { D}$, from (\ref{eq:norm}) follows  that when $A>0$ the
negative sign corresponds to the  timelike eigenvector while the  positive sign
to the  spacelike eigenvector. When  $A<0$ we have the opposite case. So  the
function $\Psi _{,z}$ determines the sign of the norm.

Let $V^a$ be the timelike eigenvector, $V_aV^a = -1$, and 
$W^a$ the spacelike  eigenvector, $W_aW^a = 1$.
In terms of the  orthonormal tetrad or comoving observer  ${{\rm e}_{\hat a}}^b
= \{ V^b , W^b , X^b , Y^b \}$,  the SEMT and the surface electric current density may be decomposed as  
\begin{subequations}\begin{eqnarray}
S_{ab} \ &=& \ \epsilon V_a V_b + p_\varphi W_a W_b , \label{eq:emtcov} \\
\mbox{\sl j}_a \ &=& \ \mbox{\sl j}^{\hat 0} V_a + 
\mbox{\sl j}^{\hat 1} W_a,  \label{eq:ja} 
\end{eqnarray}\end{subequations}
where
\begin{equation}
\epsilon \ = \ - \lambda_{\pm}, \quad \quad  p_\varphi
 \ = \ \lambda_{\mp},
\label{eq:enpr}
\end{equation}
are, respectively, the surface energy density, the  azimuthal pressure, and
\begin{equation}
\mbox{\sl j}^{\hat 0} = - V^a \mbox{\sl j}_a,  \quad 
\mbox{\sl j}^{\hat 1}  =  W^a \mbox{\sl j}_a,  \label{eq:djs}
\end{equation}
are the surface electric charge density and the azimuthal current  density  of the disk 
measured by this observer.  In (\ref{eq:enpr}) the sign is chosen according to which  is the
timelike  eigenvector and which is the spacelike eigenvector.
However, in order to satisfy  the strong energy condition
$\varrho= \epsilon + p_\varphi  \geq 0$, where $\varrho$ is
the effective Newtonian density, we must choose 
$\xi _-$ as the timelike eigenvector and  $\xi _+$ as the
spacelike eigenvector.
These condition characterizes a disk made of matter with 
the usual gravitational attractive property.
Consequently  $\Psi _{,z}$ must be taken positive.
So we have
\begin{equation}
\epsilon \ = \ - \lambda_-, \quad \quad  p_\varphi
 \ = \ \lambda_+,
\end{equation}
and
\begin{subequations}\begin{eqnarray}
V^0 &=& \frac{\nu e^{-\Psi}} { \sqrt{- 2 N_-}} (S^0_0 - S^1_1 - \sqrt D),  \\
V^1 &=&  \frac{2 \nu  e^{-\Psi}} {   \sqrt{- 2 N_-}} S^1_0 ,
\end{eqnarray}\end{subequations}
where $ \nu = \pm 1$ so that the sign is chosen  according  to  the causal
character of the timelike eigenvector  (observer's four-velocity),  
\begin{subequations}\begin{eqnarray}
W^0 &=& \frac{2} {\sqrt{2 M}}S^0_1,  \\
W^1 &=& \frac{1} { \sqrt{2 M}} (S^1_1-S^0_0 + \sqrt{ D} ),
\end{eqnarray}\end{subequations}
where
\begin{equation}
M = \sqrt{ D} \left \{ g_{11} \sqrt{ D}  +2r{\cal W} B
+ (r^2 e^{-2\Psi} + {\cal W}^2  e^{2\Psi} )A  \right \}.
\end{equation}

When $D < 0$, the eigenvalues $\lambda_\pm$ and the 
eigenvectors $\xi_\pm$ are complex  conjugates, 
 $\lambda_- = {\bar \lambda_+}$, $\xi^a_- = {\bar \xi^a_+}$.
In a comoving orthonormal tetrad ${{\rm e}_{\hat a}}^b = 
\{ V^b, W^b , X^b , Y^b \}$ the eigenvectors $\xi^a_\pm$ 
can be expressed as $\xi^a_\pm = V^a \pm i W^a$.
So we can write the SEMT in the canonical form
\begin{equation}
S_{ab} \ = \ \epsilon V_a V_b + q ( V_a W_b + W_a V_b ) 
+ p_\varphi W_a W_b,
\label{eq:emtqcov}
\end{equation}
so that it can be interpreted as the energy-momentum tensor
of a matter distribution with propagation of heat in the
tangential direction. The energy density, the azimuthal 
pressure, and the heat flow function are respectively   
\begin{equation}
\epsilon \ = \ - \ \frac{T}{2}, \quad  \quad p_\varphi 
\ = \ \frac{T}{2}, \quad  \quad q \ = \ \ \frac{\sqrt{- D}}
{2}.
\label{eq:enprq}
\end{equation}


\section{Counterrotating charged dust disks}

We now consider, based on Refs. \cite{GE} and \cite{GG2}, the possibility that
the SEMT $S^{ab}$ and the current density  $\mbox{\sl j}^a$ can be written as
the superposition of two counterrotating charged  fluids that circulate in
opposite directions; that is, we assume 
\begin{subequations}\begin{eqnarray}
S^{ab} &=& S_+^{ab} \ + \ S_-^{ab} \ , \label{eq:emtsum}   \\
\mbox{\sl j}^a    &=& \mbox{\sl j}_+^a + \mbox{\sl j}_-^a,  \label {eq:corsum} \end{eqnarray}\end{subequations}
where the  quantities on the right-hand side are, respectively, the SEMT and
the current density of the prograde and retrograde counterrotating fluids. 

Let  $U_\pm^a = ( U_\pm^0 , U_\pm^1, 0 , 0 )= U_\pm^0(1, \omega _\pm, 0 , 0 )$
be the velocity vectors of the two counterrotating fluids, where  $\omega_\pm =
U_\pm^1/U_\pm^0$ are the angular velocities of each stream.  In order to do the
decomposition  (\ref{eq:emtsum}) and (\ref{eq:corsum}) we project the velocity
vectors onto the tetrad ${{\rm e}_{\hat a}}^b$, using the relations 
\cite{CHAN}
\begin{equation}
U_\pm^{\hat a} \ = \ {{\rm e}^{\hat a}}_b U_\pm^b, \quad  \quad U_\pm^ a = \ 
{{\rm e}_{\hat b}}^a U_\pm^{\hat b} .
\end{equation}
In terms of  the tetrad (\ref{eq:tetrad}) we can write
\begin{equation}
U_\pm^a \ = \ \frac{ V^a + v_\pm W^a }{\sqrt{1 - v_\pm^2}} , \label{eq:vels}
\end{equation}
so that 
\begin{subequations}\begin{eqnarray}
&V^a &= \ \frac{\sqrt{1 - v_-^2} v_+ U_-^a - \sqrt{1 - v_+^2} v_- U_+^a}{ v_+ -
v_-}  , \label{eq:va} \\
&W^a &= \ \frac{\sqrt{1 - v_+^2} U_+^a - \sqrt{1 - v_-^2} U_-^a}{v_+ - v_-}  ,
\label{eq:wa}
\end{eqnarray}\end{subequations}
where $v_\pm = U_\pm^{\hat 1} / U_\pm^{\hat 0}$ are the tangential velocities
of the streams with respect to the tetrad.

Another quantity related with the counterrotating motion is the specific
angular momentum of a particle rotating at a radius $r$, defined as $h_\pm =
g_{\varphi a} U_\pm^a$.
This quantity can be used to analyze the stability of circular
orbits of test particles 
against radial
perturbations. The condition of stability,
\begin{equation}
\frac{d(h^2)}{dr} \ > \ 0  ,
\end{equation}
is an extension of Rayleigh criteria of stability of a fluid in rest in a
gravitational field \cite{FLU}. For an analysis of the stability
of a rotating fluid  taking into account the collective behavior of the particles see for example Refs. \cite{FHS,UL1}.

Substituting  (\ref{eq:va}) and (\ref{eq:wa}) in  (\ref{eq:emtqcov}) we obtain
\begin{eqnarray}
S^{ab} & = & \frac{ F( v_- , v_- ) (1 - v_+^2) \ U_+^a U_+^b }{(v_+ - v_-)^2}
\nonumber	\\
& + & \frac{ F( v_+ , v_+ ) (1 - v_-^2) \ U_-^a U_-^b }{(v_+ -
v_-)^2}			\nonumber	\\
& - & \frac{ F( v_+ , v_- ) (1 - v_+^2)^{\frac{1}{2}} (1 - v_-^2)^{\frac{1}{2}}
( U_+^a U_-^b + U_-^a U_+^b ) }{(v_+ - v_-)^2},		\nonumber	
\end{eqnarray}
where
\begin{equation}
F( v_1 , v_2 ) \ = \  \epsilon  v_1 v_2  - q(v_1 + v_2)  
+ p_\varphi.  \label{eq:fuu}
\end{equation}
Clearly, in order to cast the SEMT in the form (\ref{eq:emtsum}), the mixed
term must be absent and therefore the counterrotating tangential velocities
must satisfy the  following constraint
\begin{equation}
F( v_+ , v_- ) \ = \ 0  , \label{eq:liga}
\end{equation}
where we assume that $|v_\pm| \neq 1$. 

Then, assuming a given choice for the tangential velocities in agreement with
the above relation, we can write the SEMT as (\ref{eq:emtsum}) with
\begin{equation}
S^{ab} _\pm = \epsilon_\pm \ U_\pm ^a U_\pm ^b, \label{eq:sabcon}
\end{equation}
so that we have two counterrotating dust fluids with energy densities, measured in the coordinates frames, given by 
\begin{equation}
\epsilon_\pm =  \left[ \frac{ 1 - v_\pm^2 }{v_\mp - v_\pm}\right]  ( \epsilon v_ \mp - q ), \label{eq:epcon}
\end{equation}
Thus the SEMT $S^{ab}$ can be written as the superposition of two
counterrotating dust fluids if, and only if, the constraint (\ref{eq:liga}) 
admits a solution such that $v_+ \neq v_-$. 

Similarly, substituting  (\ref{eq:va}) and (\ref{eq:wa}) in  (\ref{eq:ja}) we
can write the current density as (\ref{eq:corsum}) with
\begin{equation}
\mbox{\sl j}^a_\pm  = \sigma _\pm U_\pm ^a  \label{eq:jacon}
\end{equation}
where $\sigma _\pm$ are the counterrotating electric charge densities,  measured in the coordinates frames, 
which are given by
\begin{equation}
\sigma _{ \pm} =  \left[ \frac { \sqrt{1-v^2_\pm }} 
{  v_\pm -v_\mp} \right] (\mbox{\sl j}^{\hat 1}
- \mbox{\sl j}^{\hat 0} v_\mp  ). \label{eq:sig} 
\end{equation}
Thus, we have a disk  makes of two counterrotating charged dust fluids with
energy densities given by  (\ref{eq:epcon}), and electric charge densities
given by  (\ref{eq:sig}).


As we can see from Eqs. (\ref{eq:vels}), (\ref{eq:epcon})  and (\ref{eq:sig}),
all the main physical quantities  associated with the CRM  depend on  the
counterrotating tangential velocities $v_\pm$.   However, the constraint 
(\ref{eq:liga}) does not determine  $v_\pm$ uniquely so that we need to impose 
some additional requirement in order to obtain a complete determination of  the
tangential velocities leading  to a well defined CRM.


A possibility, commonly assumed \cite{LBZ,KLE3}, is to take the two  counterrotating streams as
circulating  along electrogeodesics. Now, if the electrogeodesic equation
admits solutions corresponding to circular orbits, we can write this equation
as
\begin{equation}
\frac 12 \epsilon _\pm g_{ab,r}U^a_\pm U^b_\pm = - \sigma _\pm F_{ra} U^a_\pm.
\label{eq:elecgeo}
\end{equation}
In terms of $\omega _\pm$ we obtain
\begin{equation}
\frac 12 \epsilon _\pm (U^0_\pm)^2 ( g_{11,r} \omega _\pm ^2  + 2g_{01,r}
\omega _\pm + g_{00,r} ) = - \sigma _{\pm} U_\pm ^0 ( A _{t,r} + A_{\varphi,r}
\omega _{\pm}).  \label{eq:elecgeo1}
\end{equation}
From (\ref{eq:emtsum}), (\ref{eq:corsum}), (\ref{eq:sabcon}),
and (\ref{eq:jacon}) we have
\begin{subequations}\begin{eqnarray}
\sigma_\pm U^0_\pm & = & \frac{\mbox{\sl j}^1 -  \omega_\mp \mbox{\sl
j}^0}{\omega_\pm - \omega_\mp}, \label{eq:sigmau0} \\
\epsilon_\pm (U^0_\pm)^2 & = & \frac{S^{01} - \omega_\mp  S^{00}}{\omega_\pm -
\omega_\mp}, \label{eq:epsilonu02} \\
\omega_\mp & = &  \frac {S^{11} - \omega_\pm S^{01} } {S^{01} - \omega _ \pm
S^{00}}, \label{eq:omegapm}
\end{eqnarray}\end{subequations}
and substituting  (\ref{eq:sigmau0}) and (\ref{eq:epsilonu02}) in 
(\ref{eq:elecgeo1}) we find  
\begin{eqnarray}
\frac 12 (S^{01} - \omega_\mp S^{00}) ( g_{11,r} \omega _\pm ^2 + 2g_{01,r}
\omega _\pm + g_{00,r} ) =  \nonumber \\
 - (\mbox{\sl j}^1 -  \omega_\mp \mbox{\sl j}^0) ( A
_{t,r} + A_{\varphi,r}  \omega _{\pm}),  \label{eq:elecgeo2}
\end{eqnarray}
and using (\ref{eq:omegapm}) we obtain
\begin{eqnarray}
\frac 12[(S^{01})^2 - S^{00}S^{11}] ( g_{11,r} \omega _\pm ^2 + 2g_{01,r}
\omega _\pm + g_{00,r} )  =   \nonumber  \\
 - [S^{01}\mbox{\sl j}^1 - S^{11}\mbox{\sl j}^0    
 +  \omega_\pm (S^{01} \mbox{\sl j}^0 - S^{00}\mbox{\sl j}^1)] ( A _{t,r} +
A_{\varphi,r} \omega _{\pm}).  \label{eq:elecgeo3}
\end{eqnarray}
Therefore we conclude that 
\begin{equation}
\omega _\pm = \frac {-T_2 \pm \sqrt{T_2^2 - T_1 T_3}} {T_1}  \label{eq:omega}
\end{equation}
with
\begin{subequations}\begin{eqnarray}
T_1 &=& g_{11,r} + 2 A_{\varphi,r} \frac{\mbox{\sl j}^0  S^{01} - \mbox{\sl
j}^1 S^{00} } { S^{01}S^{01} - S^{00} S^{11} } , \label{eq:T1} \\
T_2 &=& g_{01,r} + A_{t,r} \frac{ \mbox{\sl j}^0 S^{01}  - \mbox{\sl j}^1
S^{00} } {S^{01} S^{01} - S^{00} S^{11}} + A_{\varphi,r}  \frac{ \mbox{\sl j}^1
S^{01}  - \mbox{\sl j}^0 S^{11} } { S^{01} S^{01} - S^{00} S^{11} }, 
\label{eq:T2} \\
T_3 &=& g_{00,r}  + 2 A_{t,r} \frac{\mbox{\sl j}^1 S^{01}  - \mbox{\sl j}^0
S^{11}} {S^{01} S^{01} - S^{00} S^{11} }. \label{eq:T3}
\end{eqnarray}\end{subequations}

It is easy to see  when $D>0$ that
electrogeodesic  velocities
satisfies the constraint  (\ref{eq:liga}).  
In the fact,  in terms of $\omega _\pm$ we get
\begin{equation}
v_{\pm} = - \left [ \frac{W_0 + W_{1} \omega _\pm} {V_0 + V_{1} \omega _\pm}
\right ], \label{eq:vrotcon}
\end{equation}
and, by using (\ref{eq:omega}), we have that
\begin{equation}
v_+v_- = \frac{T_1 W_0^2-2T_2W_0 W_1+ T_3 W_1^2  }{T_1 V_0^2-2T_2V_0 V_1+ T_3
V_1^2 },
\end{equation}
so that, using  (\ref{eq:emtcov}), we get
\begin{eqnarray}
F(v_+,v_-) & = &\frac{32 e^{4(\Psi -\Lambda)} \Lambda _{,z}^2  ( r^2 \Lambda 
_{,z} \sqrt{D} +4r^2 \Lambda _{,z} \Psi_{,z} e^{\Psi - \Lambda} - {\cal
W}_{,z}^2 e^{5\Psi - \Lambda}  ) } {r^3(A+ \sqrt{D}) p_\varphi (S^{01}S^{01} -
S^{00} S^{11})(T_1 V_0^2-2T_2 V_0 V_1 +T_3V_1^2)} \nonumber   \\
& & \times \left [ \Lambda _{,z} - 2r \Psi_{,r} \Psi_{,z}
 + \frac{1}{2r} {\cal
W}_{,r} {\cal W}_{,z} e^{4 \Psi} 
+\frac{1}{r}(r^2e^{-2 \Psi} -{\cal W}^2 e^{2
\Psi}) A_{t,r} A_{t,z}  \right . \nonumber \\  
&& \left . \ - \frac{1}{r}A_{\varphi,r} A_{\varphi,z} e^{2 \Psi} 
+ \frac{1}{r}{\cal W}(A_{\varphi,r} A_{t,z}+ A_{\varphi,z}
A_{t,r})e^{2 \Psi}  \right ]. 
\label{eq:liga3}\end{eqnarray}
Finally, using the Einstein-Maxwell equation (\ref{eq:lamz}) follows
immediately that $F(v_+,v_-)$ vanishes and therefore the electrogeodesic
velocities satisfy the  constraint (\ref{eq:liga}) and so, if the
electrogeodesic equation admits solutions corresponding to circular orbits, we
have a well defined CRM.  
When there is   heat flow a proof as the previous one 
is not trivial  and in this case  we take   the 
counterrotating  hypothesis (\ref{eq:sabcon})  and 
(\ref{eq:jacon})  as an ansatz.  


Another  possibility is to take the two
counterrotating fluids not circulating along  electrogeodesics but with equal
and opposite tangential  velocities,
\begin{equation}
v_{\pm} = \pm v = \pm \sqrt{p_\varphi / \epsilon}.
\end{equation}
This choice, that imply the existence of additional  interactions between  the
two streams (e.g. collisions), leads to a complete  determination of the
velocity vectors.  However, this can be made only when 
$0 \leq p_\varphi/ \epsilon \leq 1$. So in precense of  heat flow   by Eq. (\ref{eq:enprq}) we have 
$p_\varphi/ \epsilon =-1$  
which corresponds to an imaginary velocity and therefore
these disks have an unphysical behavior in the regions
where there is  heat flow.  

In the general case, the two counterrotating streams
circulate with different velocities and we can write (\ref{eq:liga}) as 
\begin{equation}
 v_+ v_- = - \frac {p_\varphi}{\epsilon}.
\end{equation}
However, this relation does not determine completely the tangential velocities,
and therefore the CRM is undetermined.
In summary, the counterrotating tangential velocities can be explicitly
determined only if we assume some additional relationship between them, like
the equal and  opposite condition or the electro-geodesic condition. Now, can
happen that the obtained  solutions do not satisfy any of these two conditions.
That is, the counterrotating velocities are, in general, not completely
determined by the constraint (\ref{eq:liga}). Thus, the CRM is in general
undetermined since the counterrotaing energy densities and pressures can not be
explicitly written without a knowledge of the counterrotating tangential
velocities.


\section{Disks from a Charged and magnetized Kerr-NUT 
solution}

As an example of the above presented formalism, we consider  thin disk models obtained by means of the ``displace, cut and reflect'' method applied to the electromagnetic 
generalization of the Kerr-NUT metric,
which can be written as
\begin{subequations}\begin{eqnarray}
\Psi  &=&  \frac 12 \ln \left[ \frac{a_1^2 x^2 + b_1^2 y^2 
- c^2}{u^2 + v^2} \right] , \\
\Lambda &=&  \frac 12 \ln \left [ \frac {a_1^2x^2+b_1^2y^2 -c^2}{a_1^2(x^2-y^2)} \right ] , \\
{\cal W} &=&  \frac{2 k c}{a_1} \left\{ \frac{b_1(1-y^2)}
{(a_1^2x^2+b_1^2y^2-c^2)} [ a_1 a_2 x +b_1 b_2 y  
 + \frac 1 2 c(1+c^2) ] + b_2 y \right \}, \\
A_t &=& \sqrt{ 2(c^2-1) } \left [ \frac {a_2u+b_2v}{u^2+v^2}\right],     \\
A_\varphi & = & -k \frac { \sqrt{ 2(c^2-1) } }{a_1} 
\left \{ \frac {(a_2u+b_2v)[-y(b_1y+2cb_2)+b_1]}{u^2+v^2}
+ b_2 y  \right \},
\end{eqnarray}\end{subequations}
where 
\begin{equation}
u = a_1 x + c a_2,   \ \ \ \
v = b_1 y + c b_2,
\end{equation}
and $a_1^2+b_1^2=a_2^2+b_2^2=c^2 \geq 1 $,  being $c$ the parameter that 
controls the electromagmetic field.  $x$ and $y$ are the 
prolate  spheroidal coordinates  which are related 
with the Weyl  coordinates by 
\begin{equation}
r^2   = k^2 (x^2-1)(1-y^2),  \quad  \quad z + z_0 = kxy,  \label{eq:coorp}
\end{equation}
where  $1 \leq x \leq \infty$,  $0  \leq y \leq 1 $, and $k$ is an arbitrary
constant. Note that we have displaced the origin of the $z$ axis in $z_0$.  
This solution can be generated, in these coordinates,  using  
the well-known  complex potential formalism proposed by 
Ernst \cite{E2} using as seed solution the   Kerr-NUT vacuum solution 
\cite{KSHM}.  So  when $c=1$ this solution reduces 
to the Kerr-NUT vacuum solution. When $b_1=0$ we have a
charged and magnetized Taub-NUT solution \cite{GG3}, and when  $b_2=0$
we have the well known Kerr-Newman solution. 
Let $\tilde T =kT$, $\tilde D = k^2 D$,  $\tilde 
{\mbox{\sl  j}}_t=k\mbox{\sl  j}_t$ be,  therefore
\begin{subequations}\begin{eqnarray}
\tilde T &=&  \frac{ 4ca_1 }{ (\bar x^2-\bar y^2)^{3/2}(u^2+v^2)^{3/2} }
\left \{  -a_1a_2\bar y \bar x^4 +[-2c\bar y^3- 3b_1b_2\bar y^2   \right .   \nonumber  \\ 
& & \left .   + c(1-c^2)\bar y + b_1b_2] \bar x^3  
+ 3 a_1a_2 \bar y (1-\bar y^2)\bar x^2 \right.  \nonumber \\
&& \left . + [-b_1b_2\bar y^3 
+c(1-c^2)\bar y^2+3b_1b_2\bar y+2c^3]\bar x \bar y +a_1a_2 \bar y^3 \right \},  \\
\tilde D &=&  \frac{16 c^2 a_1^2}{(\bar x^2-\bar y^2)(u^2+v^3)^3} \left \{ b_1^2(c^2\bar x^2-a_2^2)\bar y^4+2b_1b_2(c^3\bar x+c\bar x+2a_1a_2)\bar x \bar y^3 \right .  \nonumber \\
& & \left . +[c^2a_1^2\bar x^4+2a_1a_2c(c^2+1)(\bar x^2+1)\bar x -(3c^2a_1^2+3c^2a_2^2-6a_1^2a_2^2-c^6
\right . \nonumber \\
& &  \left . -4c^4-c^2)\bar x^2 +c^2a_2^2]\bar y^2  
+ 2b_1b_2(2a_1a_2\bar x+c^3+c)\bar x^2 \bar y-b_2^2(a_1^2
\bar x^2-c^2)\bar x^2  \right  \},  \\
\tilde {\mbox{\sl  j}}_t &=& - \frac{2 \sqrt{2(c^2-1)} a_1 }{(\bar x^2-\bar y^2)^{1/2}(u^2+v^2)
^{5/2}} \left \{ b_1^3b_2\bar x \bar y^4+b_1^2(3a_1a_2\bar x^2+2c^3\bar x-a_1a_2)\bar y^3 
\right . \nonumber \\
& & \left .
-b_1b_2(3a_1^2\bar x^2-3a_1^2-c^4+c^2)\bar x \bar y^2
+[-a_1^2(a_1a_2\bar x+2c^3)\bar x^3 
\right .  \nonumber   \\ 
&& \left . -a_1a_2(-3a_1^2+2c^2+c^4)\bar x^2-2c^3(c^2-2a_1^2)\bar x  +c^4a_1a_2]\bar y
\right .  \nonumber \\
& & \left .  + b_1b_2\bar x(a_1^2\bar x^2-c^4)  \right \},  \\
\mbox{\sl  j}_\varphi & = &  
\frac{2 \sqrt{2(c^2-1)} }{(\bar x^2-\bar y^2)^{1/2}(u^2+v^2)
^{5/2}}  
\left \{ -a_1a_2b_1^3(\bar x^2-1)\bar y^5
\right. \nonumber \\
&& \left . +b_1^2b_2(3a_1^2\bar x^3
-3a_1^2\bar x+c^4\bar x+c^2\bar x+2ca_1a_2)\bar y^4  
\right . \nonumber  \\
&& \left .
+b_1[a_1^2(3a_1a_2\bar x+8c^3)\bar x^3-a_1a_2(4a_1^2-7c^4-3c^2)\bar x^2
\right . \nonumber  \\
&&  \left.
+2c(-4c^2a_1^2+c^4+2a_1^2a_2^2+c^6)
\bar x-a_1a_2(b_1^2+c^4)]\bar y^3
\right . \nonumber \\
&&  \left .
-b_2[a_1^4\bar x^5-4a_1^2(c^4-b_1^2)\bar x^3-2c^3a_1a_2(2c^2-1)\bar x^2
\right . \nonumber \\
&&  \left .
+(-4c^2a_1^2+c^4+4c^4a_1^2+3a_1^4-c^8)\bar x+2c^5a_1a_2]
\bar y^2
\right . \nonumber \\
&&  \left . 
-b_1[3a_1^3a_2\bar x^4+4ca_1^2(a_2^2+c^2)\bar x^3-a_1a_2(3a_1^2-2c^2-7c^4)\bar x^2
\right . \nonumber \\
&&  \left .
-2c^3(2a_1^2-c^4-c^2)\bar x-c^4a_1a_2]\bar y 
\right . \nonumber \\
&&
\left . +b_2\bar x(a_1^2\bar x^2-c^4)(a_1^2\bar x^2+2ca_2a_1\bar x+b_1^2+c^4)
\right \} .
\end{eqnarray}\end{subequations}
In the above expressions $\bar{x}$ and  $\bar{y}$ are given by
\begin{subequations}\begin{eqnarray}
2\bar{x}  & = \sqrt {\tilde{r}^2 + (\alpha + 1)^2} + \sqrt {\tilde{r}^2 +
(\alpha - 1)^2}, \label{eq:xbar} \\
2\bar{y}  & = \sqrt {\tilde{r}^2 + (\alpha + 1)^2} - \sqrt  {\tilde{r}^2 +
(\alpha - 1)^2},\label{eq:ybar}
\end{eqnarray}\end{subequations}
where $\tilde{r}=r/k$ and $\alpha = z_0/k$, with  $\alpha >1$.

When $b_2=0$ we have  
\begin{equation}
\tilde D_{ {\rm KN} } = \frac {16 c^4 a^2 \bar y^2 \{ [a(\bar x^2 +1)+ \bar x(1+c^2)]^2 - b^2 \tilde r^2 \}} {(\bar x^2 - \bar
y^2)[(a\bar x+c^2)^2 + b^2 \bar y^2]^3}. 
\end{equation}
In order to analyze the behavior of $\tilde D_{ {\rm KN} }$,
is enough to consider the expression
\begin{equation}
 \tilde D_0 =  [a(\bar x^2 +1)+ \bar x(1+c^2)]^2 - b^2 \tilde r^2,  
\end{equation}
that can be written as
\begin{eqnarray} 
\tilde D_0 &=& a(1+c^2)R_+ [\alpha(\alpha-1)+ 2 + \tilde r ] + a(1+c^2)R_-
[\alpha(\alpha+1) + 2 + \tilde r ]  \nonumber  \\  
&& + \frac 12 R_+ R_- [(c^2+1)^2 + a^2( \tilde r^2 +  \alpha ^2+3)]   + \frac
12 [(c^2+1)^2 + 5a^2]    \nonumber \\ 
&&+\frac 12 \tilde r^2 [c^4+1+a^2 (\tilde r^2+2\alpha^2+6)] +\frac 12
\alpha^2[(c^2+1)^2 + a^2 (\alpha^2 + 2 )] ,   \nonumber \\
&&      
\label{eq:d0}
\end{eqnarray}
where $R_\pm = \sqrt {\tilde{r}^2 + (\alpha \pm 1)^2}$. Since $\alpha(\alpha \mp 1)+ 2 >0$ for any $\alpha$, from (\ref{eq:d0}) follows that $D$  always is  a positive quantity for
Kerr-Newman fields and therefore the eigenvalues of the energy-momentum tensor
are always real quantities. So we conclude that these disks can be interpreted,
for all the values of parameters, as a matter distribution with currents and
purely azimuthal pressure and without heat flow \cite{GG4}. 

\begin{figure*}
$$
\begin{array}{ccc}
\tilde D  &   \tilde D & \\ 
\epsfig{width=1.5in,file=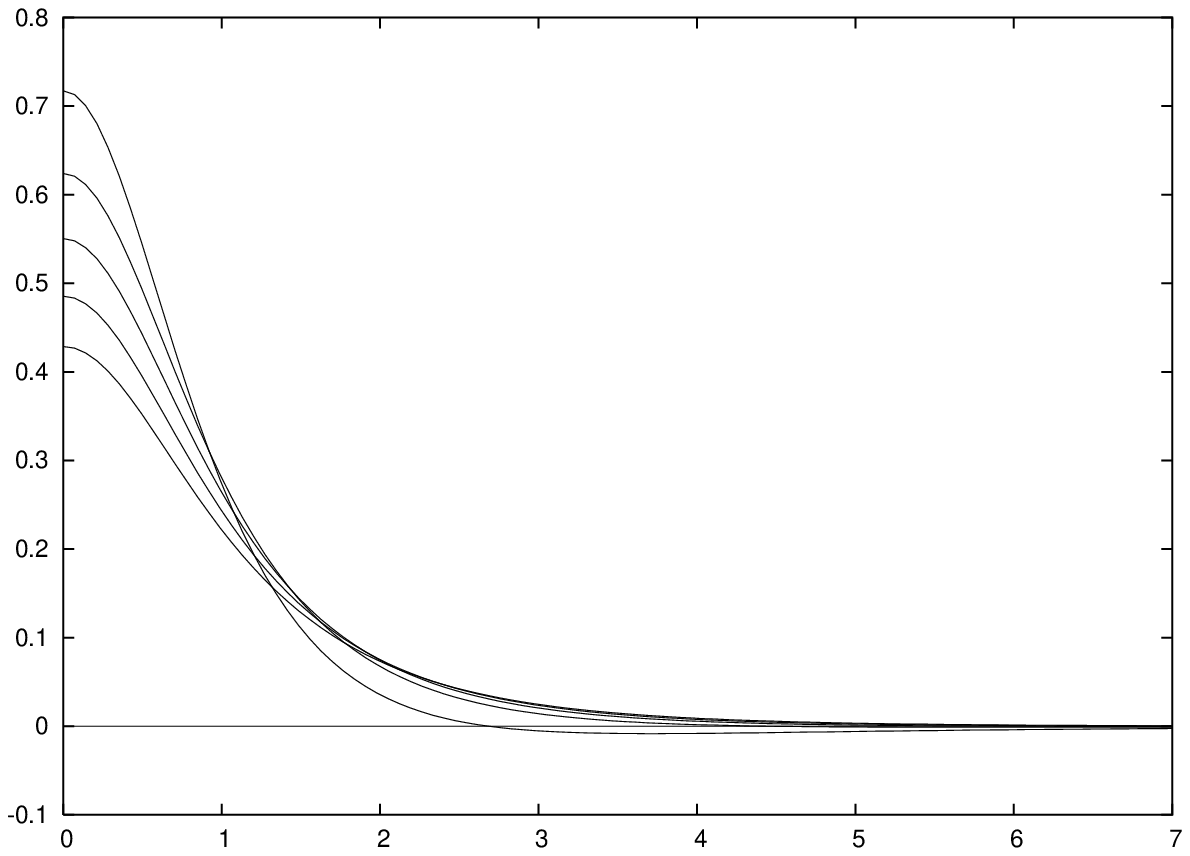} &
\epsfig{width=1.5in,file=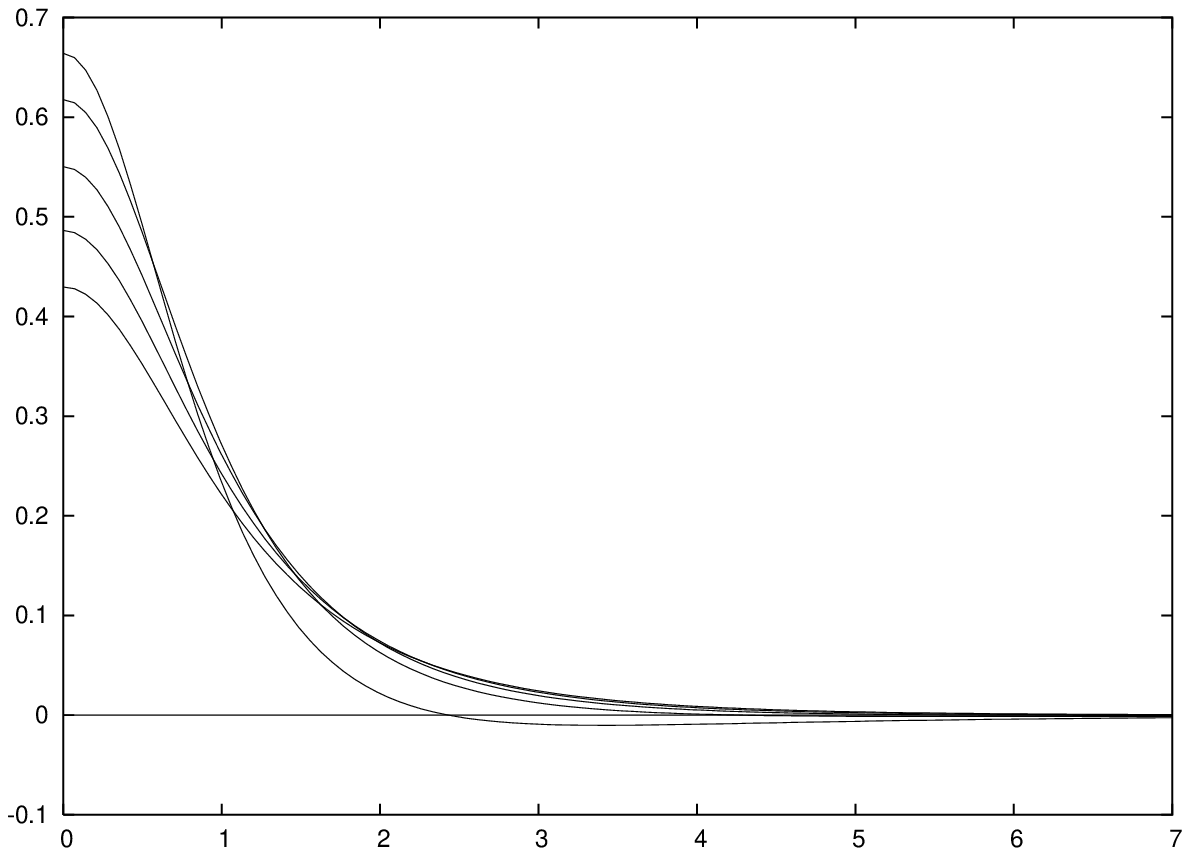} &  \\
\tilde r & \tilde r  \\
(a) & (b)   \\
\tilde \epsilon  &   \tilde \epsilon & \tilde \epsilon\\ 
\epsfig{width=1.5in,file=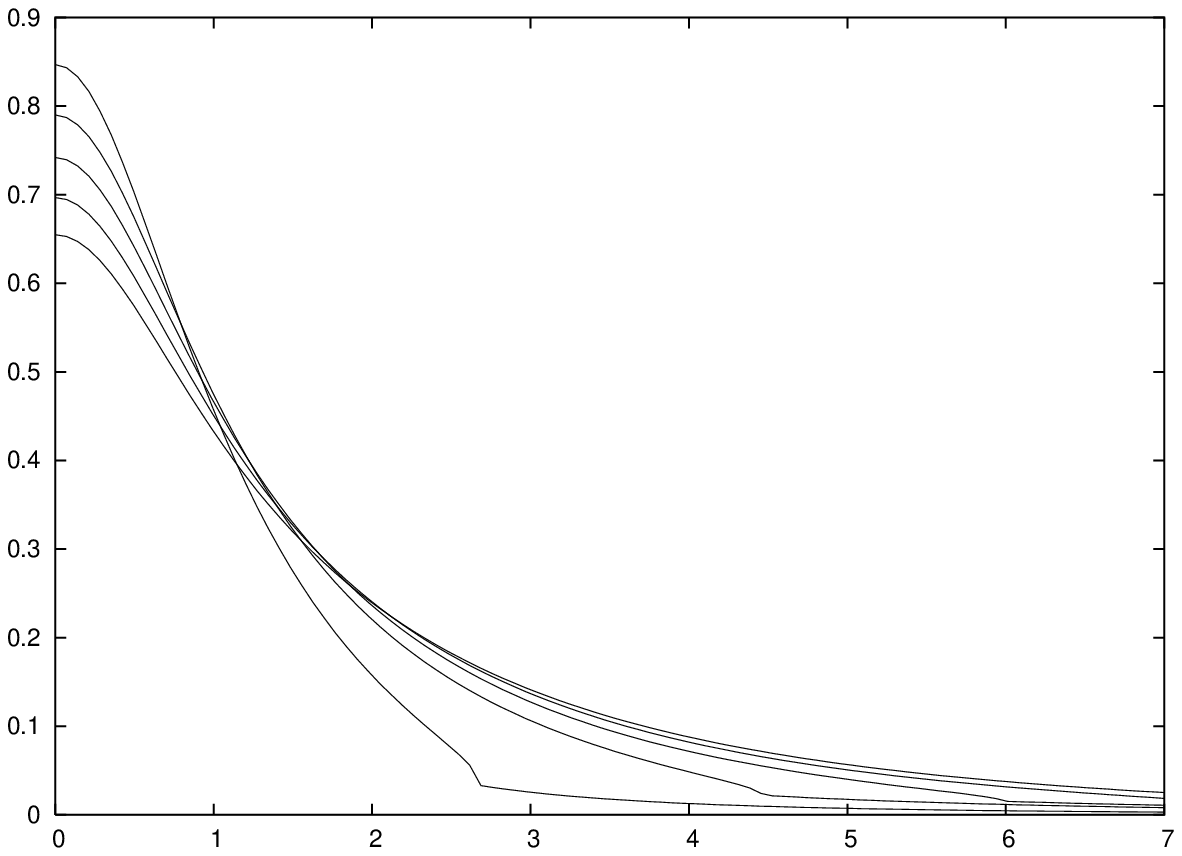} &
\epsfig{width=1.5in,file=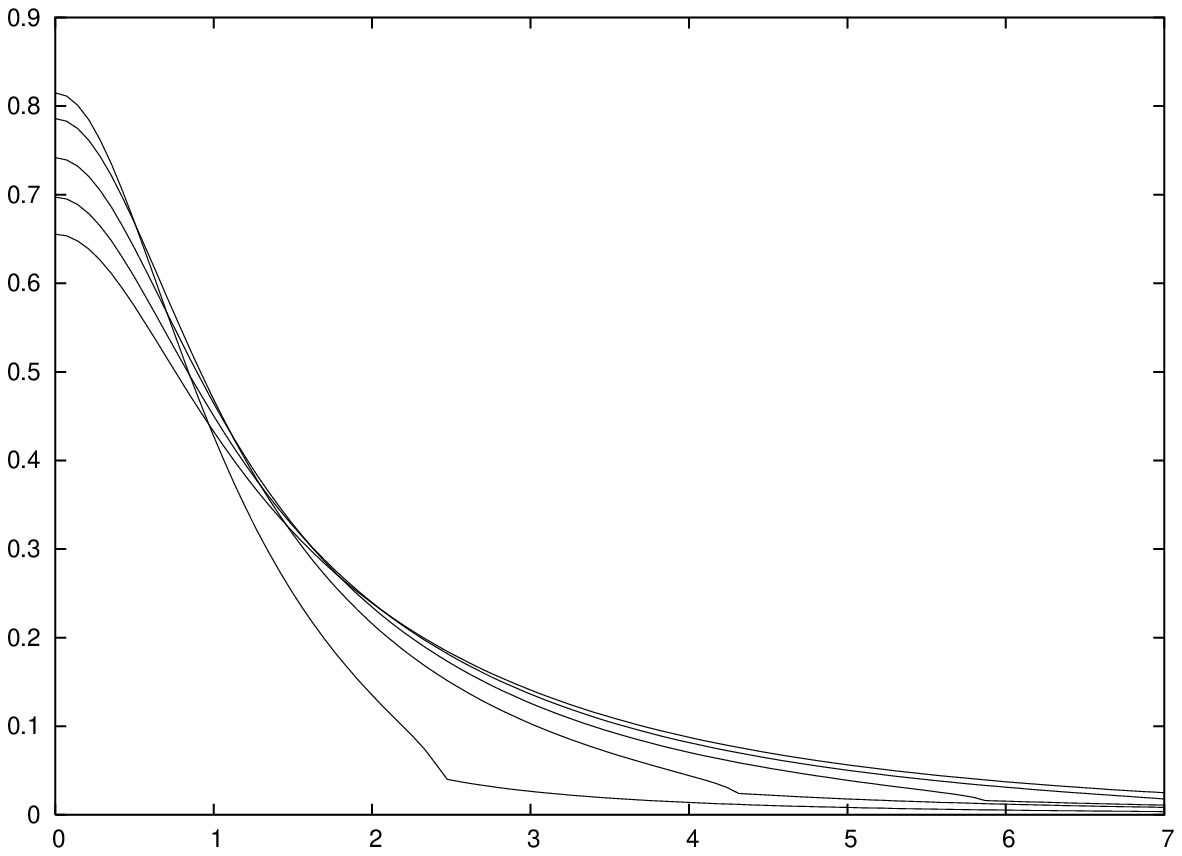} &
\epsfig{width=1.5in,file=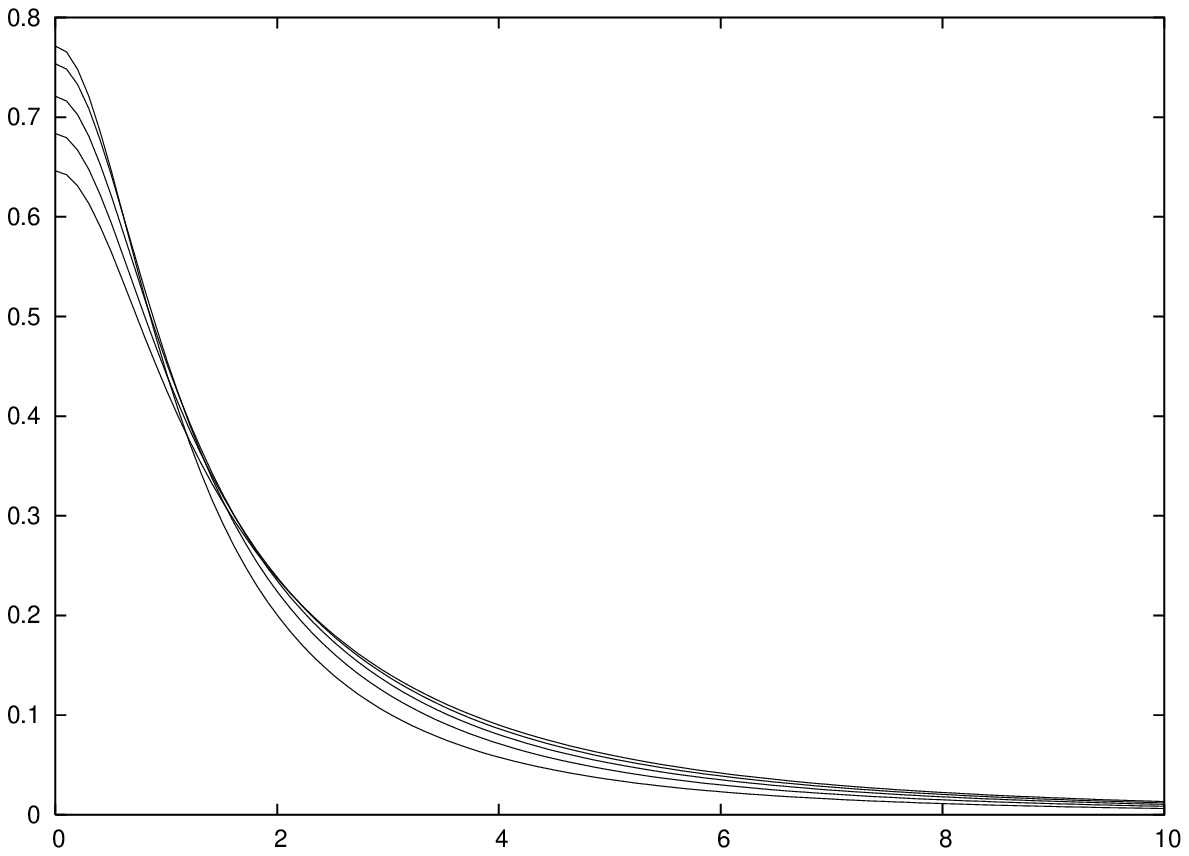}  \\
\tilde r & \tilde r & \tilde r \\
(c) & (d) & (e) \\
\tilde p_\varphi  &  \tilde p_\varphi & \tilde p_\varphi \\ 
\epsfig{width=1.5in,file=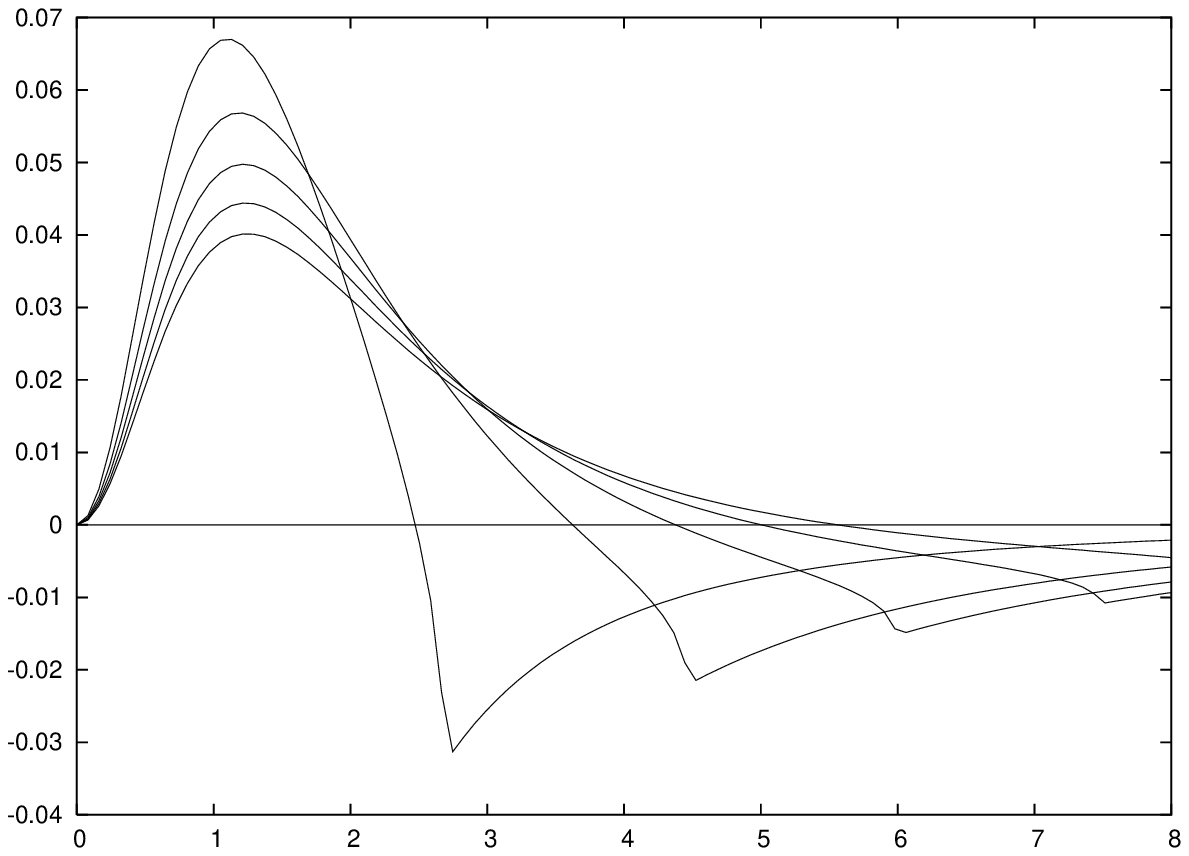} &
\epsfig{width=1.5in,file=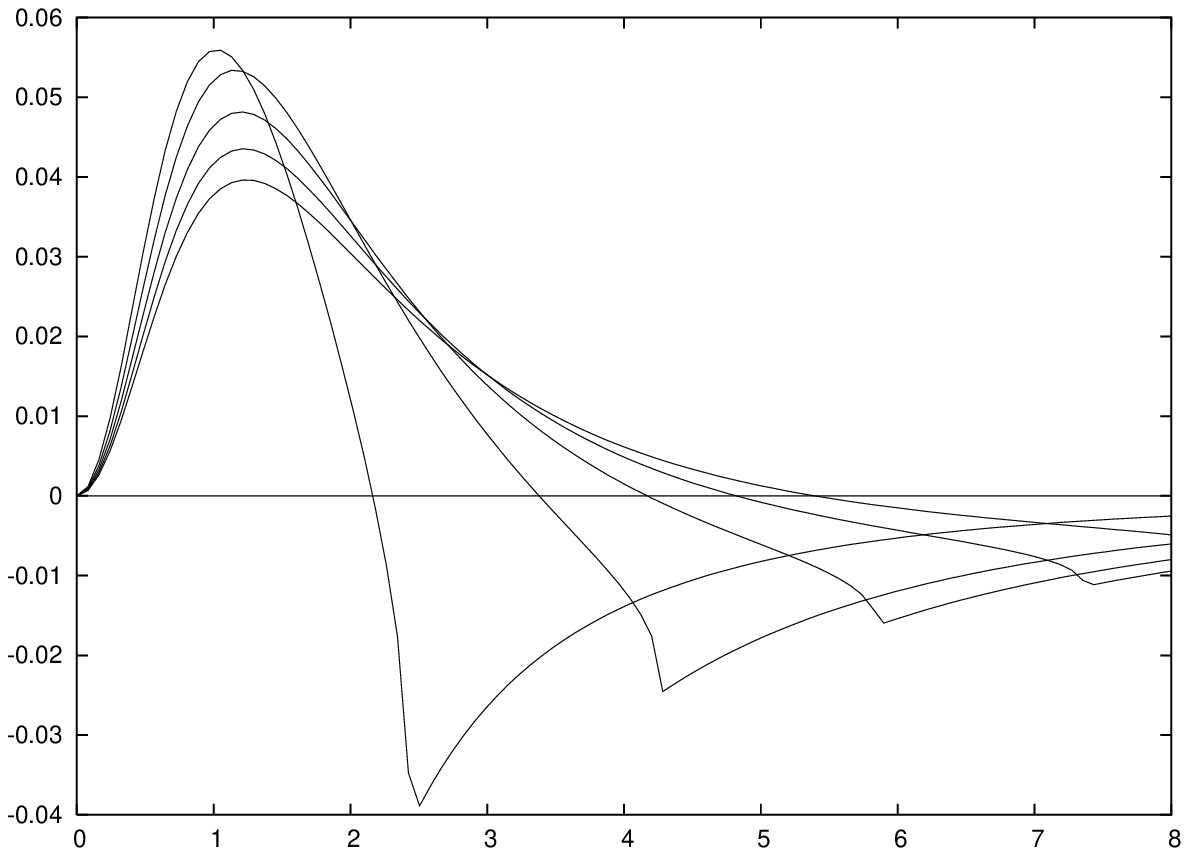} &
\epsfig{width=1.5in,file=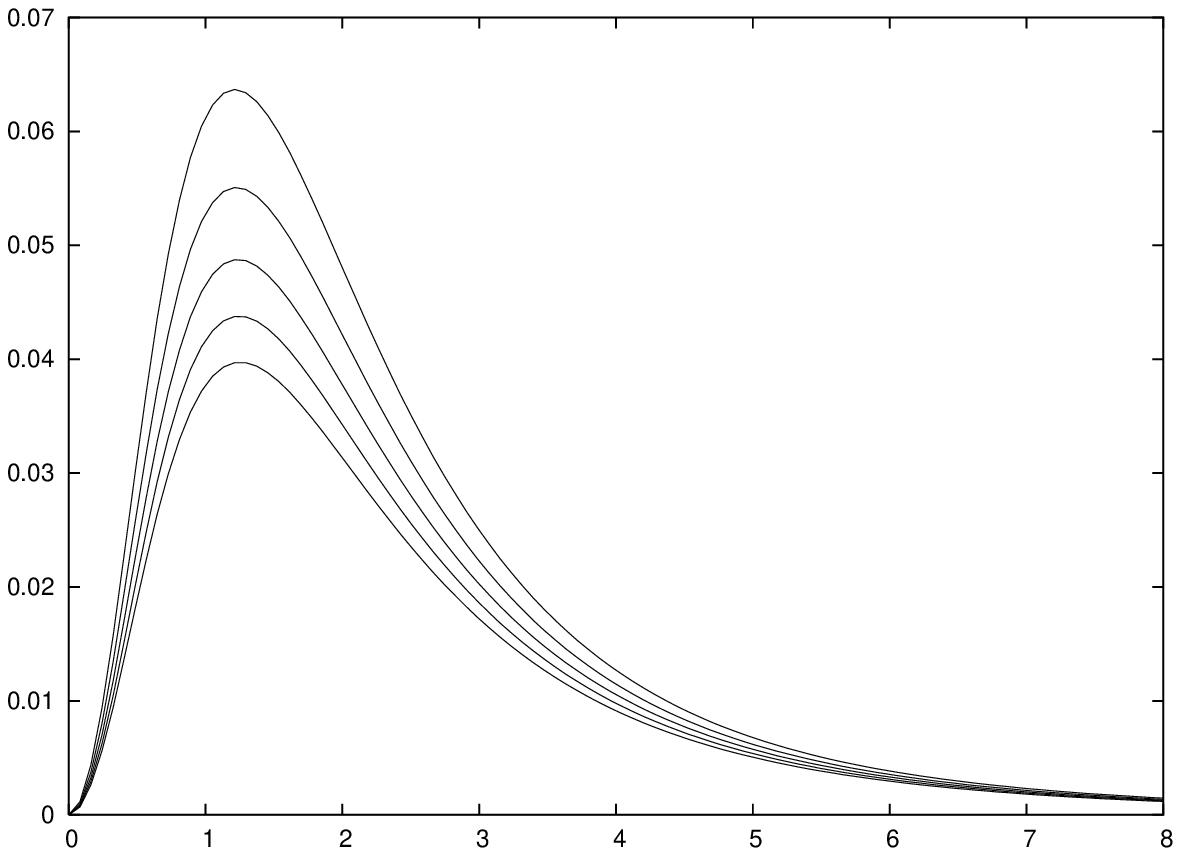}  \\
\tilde r & \tilde r & \tilde r \\
(f) & (g) & (h) \\
\tilde q  &  \tilde q &  \\ 
\epsfig{width=1.5in,file=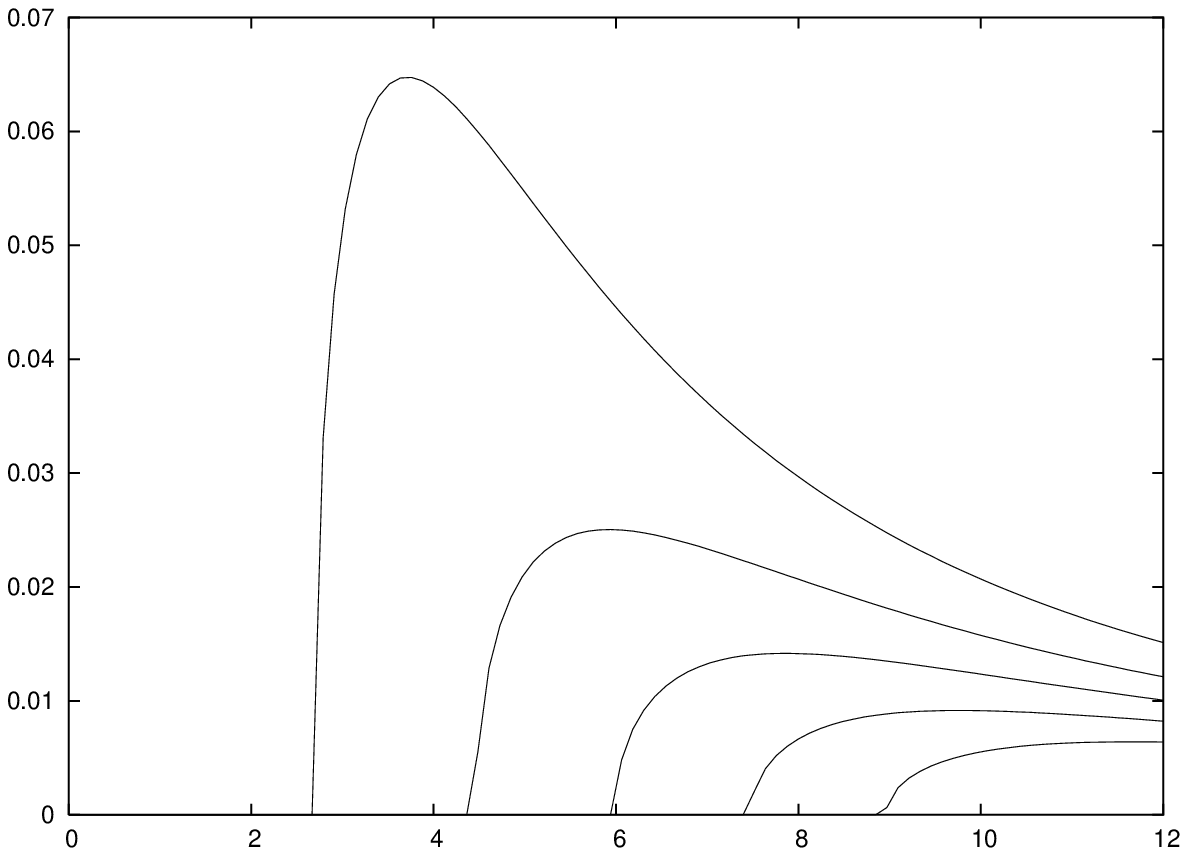} &
\epsfig{width=1.5in,file=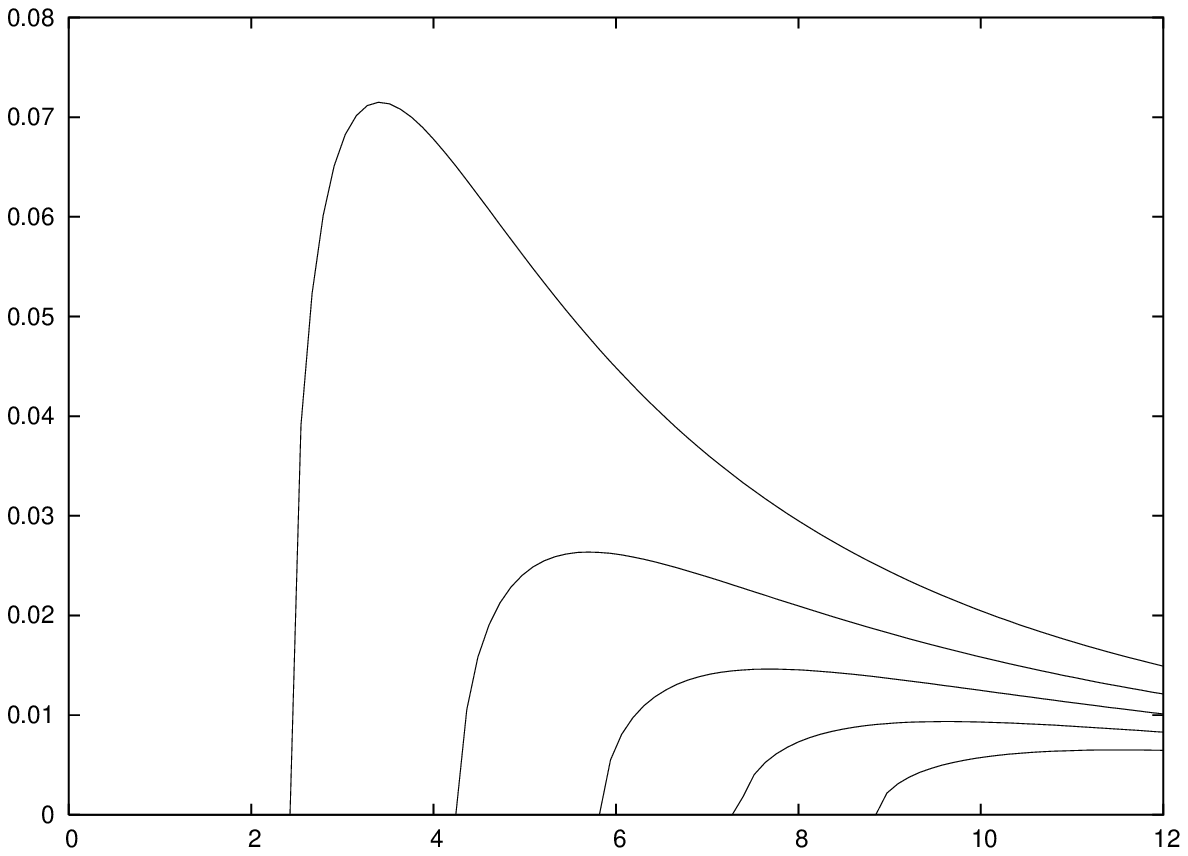} &
     \\
\tilde r & \tilde r &  \\
(i) & (j) & 
\end{array}
$$
\caption{ The discriminant $\tilde D$ $(a)$ for charged and magnetized Kerr-NUT disks  with  $b_1=0.2$, $b_2=0.9$ and 
$(b)$ for charged and magnetized Taub-NUT disks  
with  $b_2=0.9$, the energy  density $\tilde \epsilon$ 
$(c)$ for charged and magnetized Kerr-NUT disks  with  $b_1=0.2$, $b_2=0.9$, $(d)$ for charged and magnetized Taub-NUT disks  
with  $b_2=0.9$ and $(e)$ for Kerr-Newman disks with $b_1=0.2$,  the azimuthal pressure $\tilde p_\varphi$ 
$(f)$ for charged and magnetized Kerr-NUT disks  with  $b_1=0.2$, $b_2=0.9$, $(g)$ for charged and magnetized Taub-NUT disks  
with  $b_2=0.9$ and $(h)$ for Kerr-Newman disks with $b_1=0.2$, 
the  heat flow function  $\tilde q = k q$ 
$(i)$ for charged and magnetized Kerr-NUT disks  with  $b_1=0.2$, $b_2=0.9$ and  $(j)$ for charged and magnetized Taub-NUT disks  with  $b_2=0.9$, and 
$\alpha = 2.0$,  $c=1.0$
(top curves), $1.5$,  $2.0$, $2.5$, and  $3.0$ (bottom
curves), as functions of  $\tilde r$.}
\label{fig:disenprq}
\end{figure*}

In order to study the behavior of $D$ when $b_2 \neq 0$ 
and  of the other  physical  quantities associated with
the disks, we shall perform a graphical analysis of them for  
charged and magnetized Kerr-NUT disks  with 
$b_1=0.2$, $b_2=0.9$,  charged and magnetized Taub-NUT disks with  $b_2=0.9$, and Kerr-Newman disks with $b_1=0.2$, for 
$\alpha = 2$ and  $c=1.0$, $1.5$,  $2.0$, $2.5$, $3.0$, 
as functions of $\tilde r$.
In Fig. \ref{fig:disenprq}$(a)$ we plot $\tilde D$ for charged and magnetized Kerr-NUT disks 
and in Fig. 
\ref{fig:disenprq}$(b)$ we plot $\tilde D$ for charged and magnetized Taub-NUT disks. We found that $\tilde D$ takes 
negative values after certain value of $\tilde r = \tilde 
r_0$. Therefore these  disks
always have heat flow beginning at the root $\tilde r_0$.
We also computed 
this function  for other values of the parameters and, in  all the cases, we
found  the same behavior.    

In Figs. \ref{fig:disenprq}$(c)$ - 
\ref{fig:disenprq}$(h)$ we  show the energy density  $\tilde \epsilon$ and the azimuthal pressure
$\tilde p_\varphi$. We see  that the energy density is in all cases  always a positive quantity in agreement with the weak energy
condition, whereas the pressure becomes negative for a 
value of $\tilde r > \tilde r_0$ in the case of charged and
magnetized Kerr-NUT and Taub-NUT disks.  We also see that  the presence of electromagnetic  field decreases
the  energy density  at the central region of the disk and  later increases it. The heat flow function $\tilde q = kq$
is also represented in Figs. \ref{fig:disenprq}$(i)$ and
\ref{fig:disenprq}$(j)$  for charged and
magnetized Kerr-NUT and Taub-NUT disks.

\begin{figure*}
$$
\begin{array}{ccc}
\tilde { \mbox{\sl j} }_t & \tilde { \mbox{\sl j} }_t & 
\tilde { \mbox{\sl j} }_t           \\ 
\epsfig{width=2.0in,file=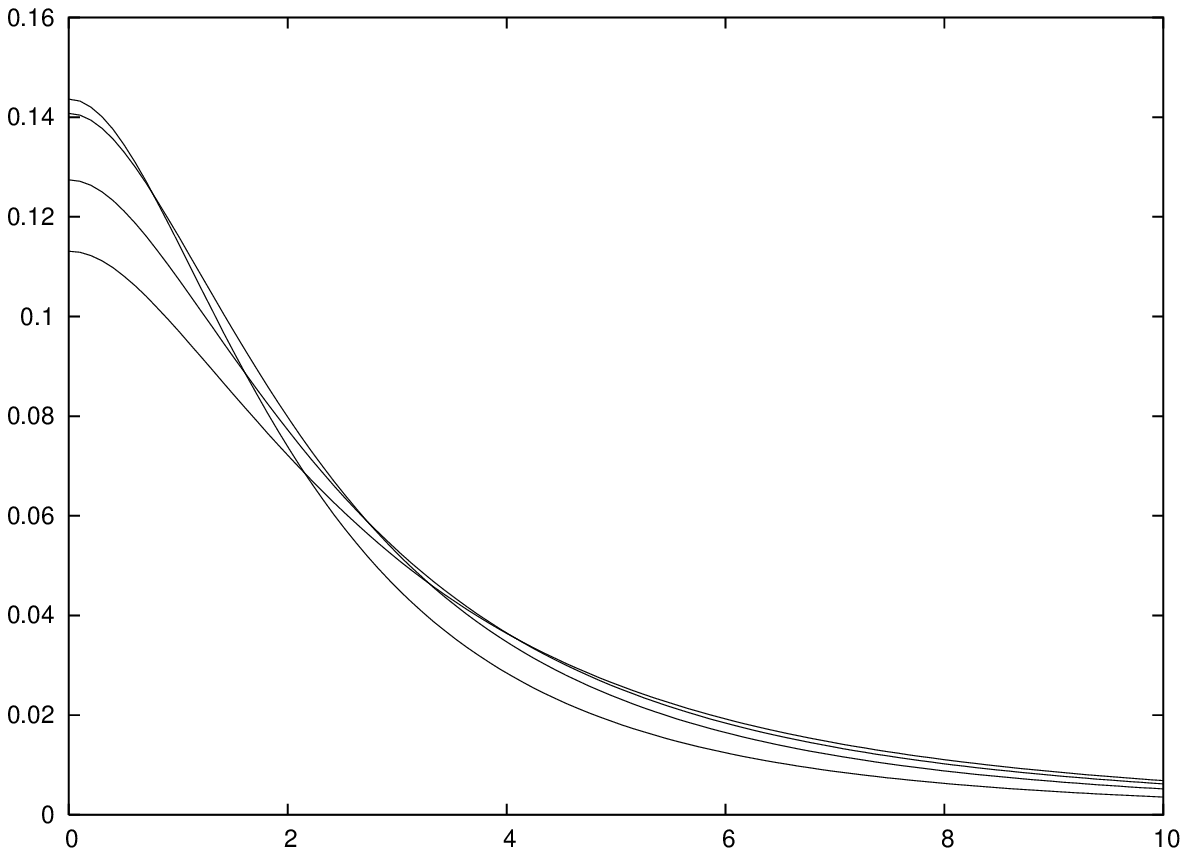} &
\epsfig{width=2.0in,file=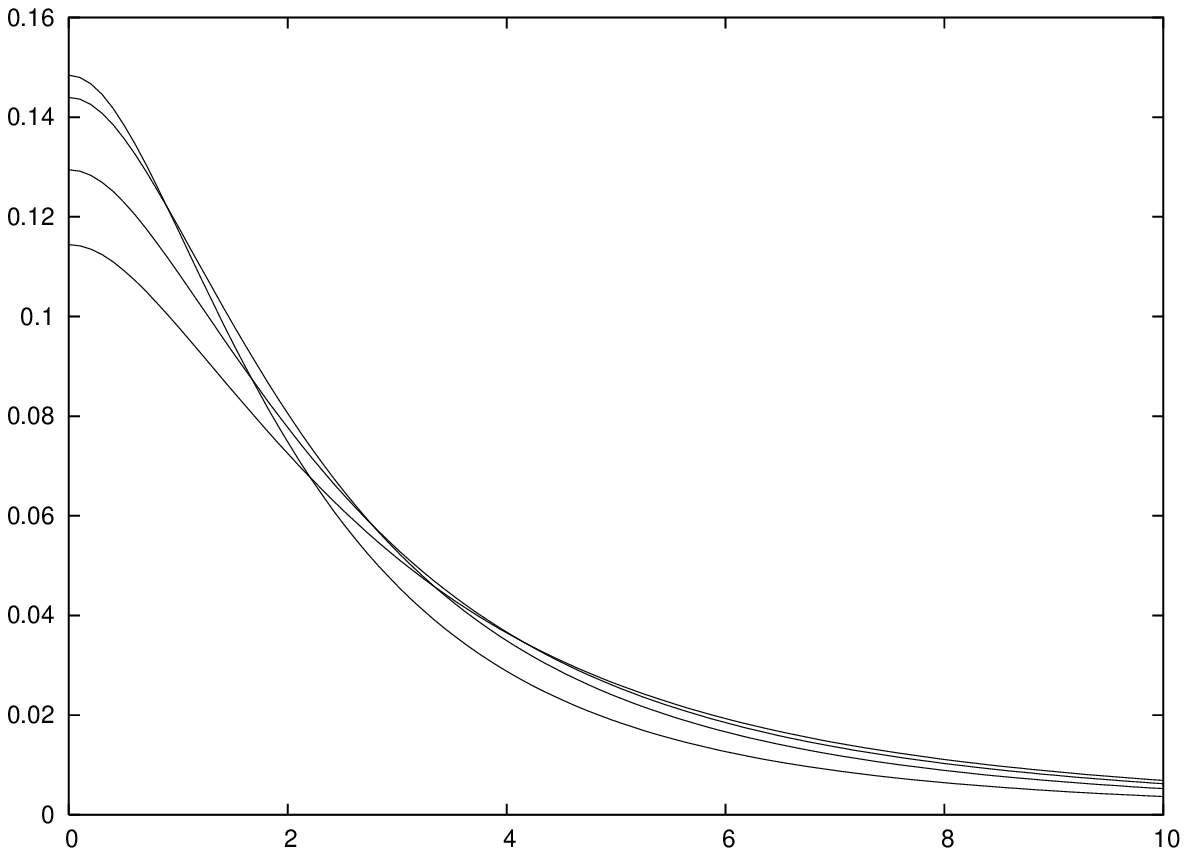} &
\epsfig{width=2.0in,file=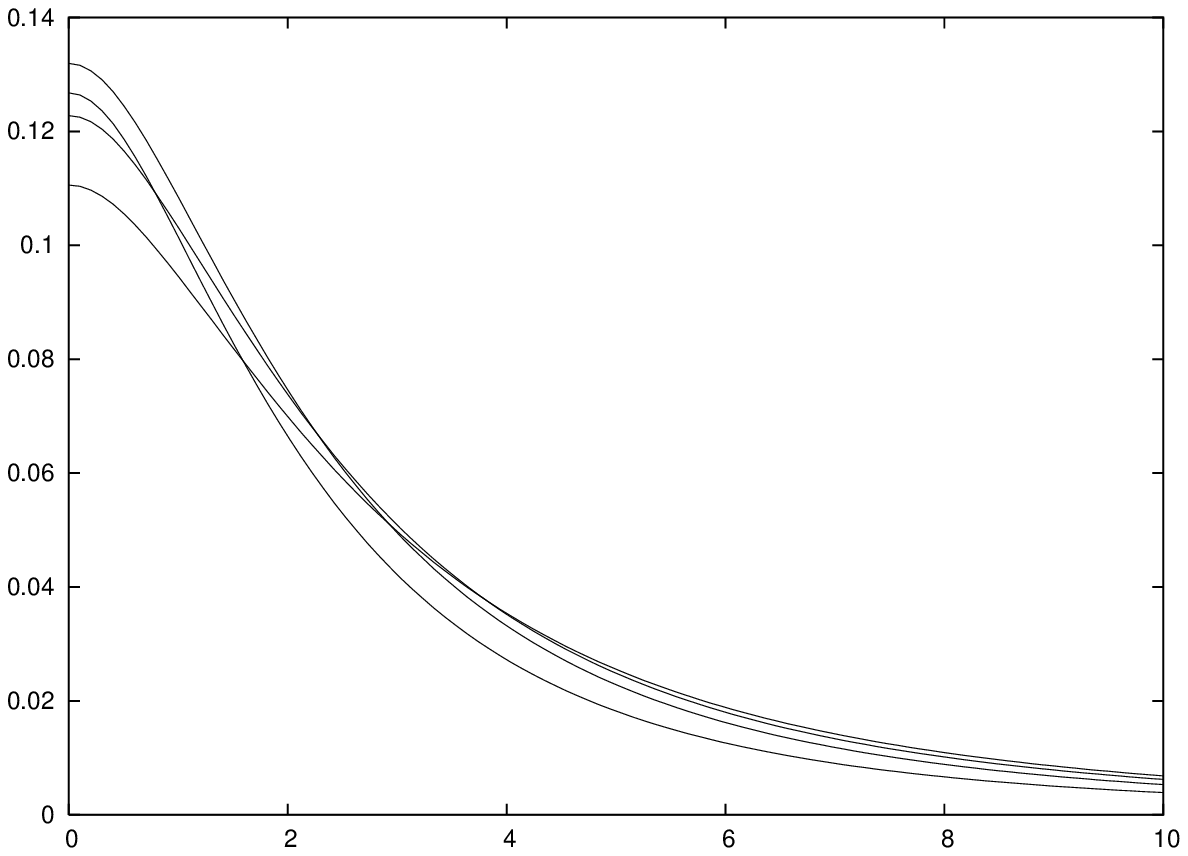}  \\
\tilde r & \tilde r & \tilde r \\
(a) & (b) & (c)  \\
\mbox{\sl j}_\varphi & \mbox{\sl j}_\varphi & 
-\mbox{\sl j}_\varphi           \\ 
\epsfig{width=2.0in,file=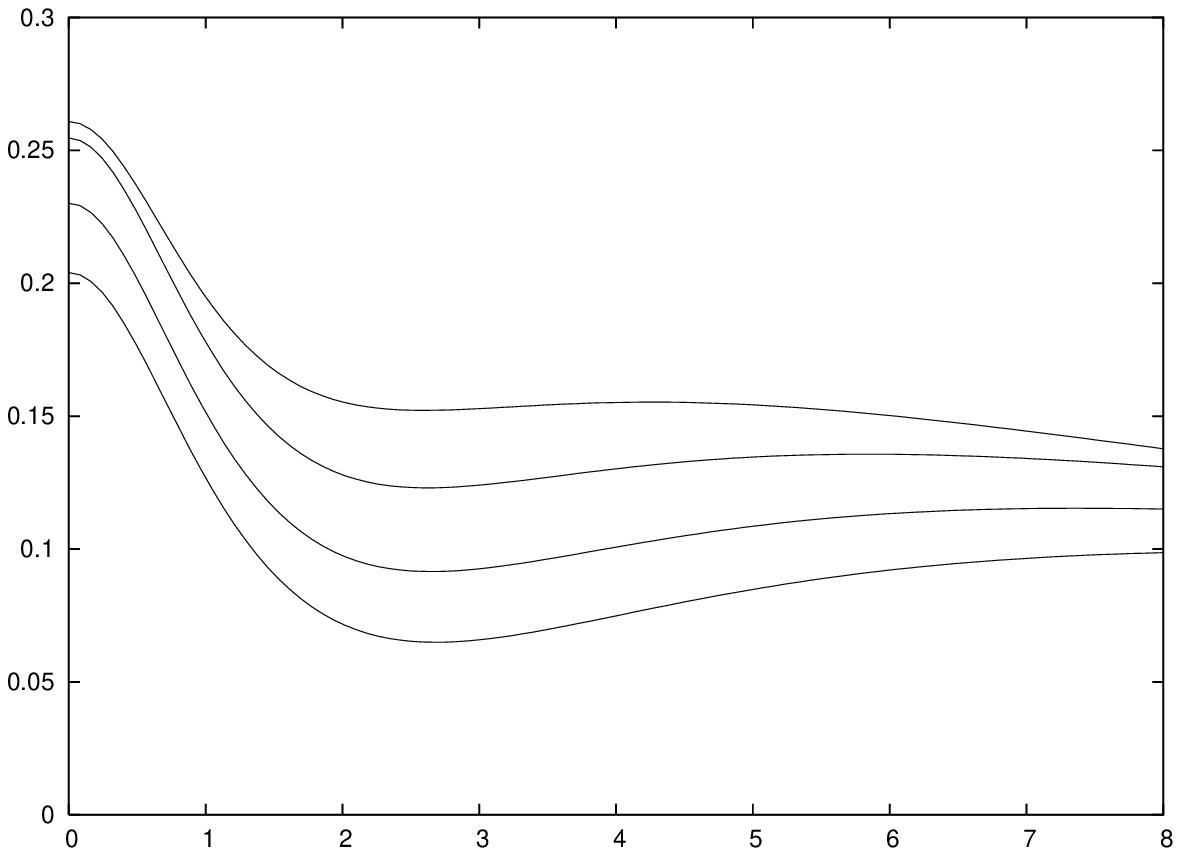} &
\epsfig{width=2.0in,file=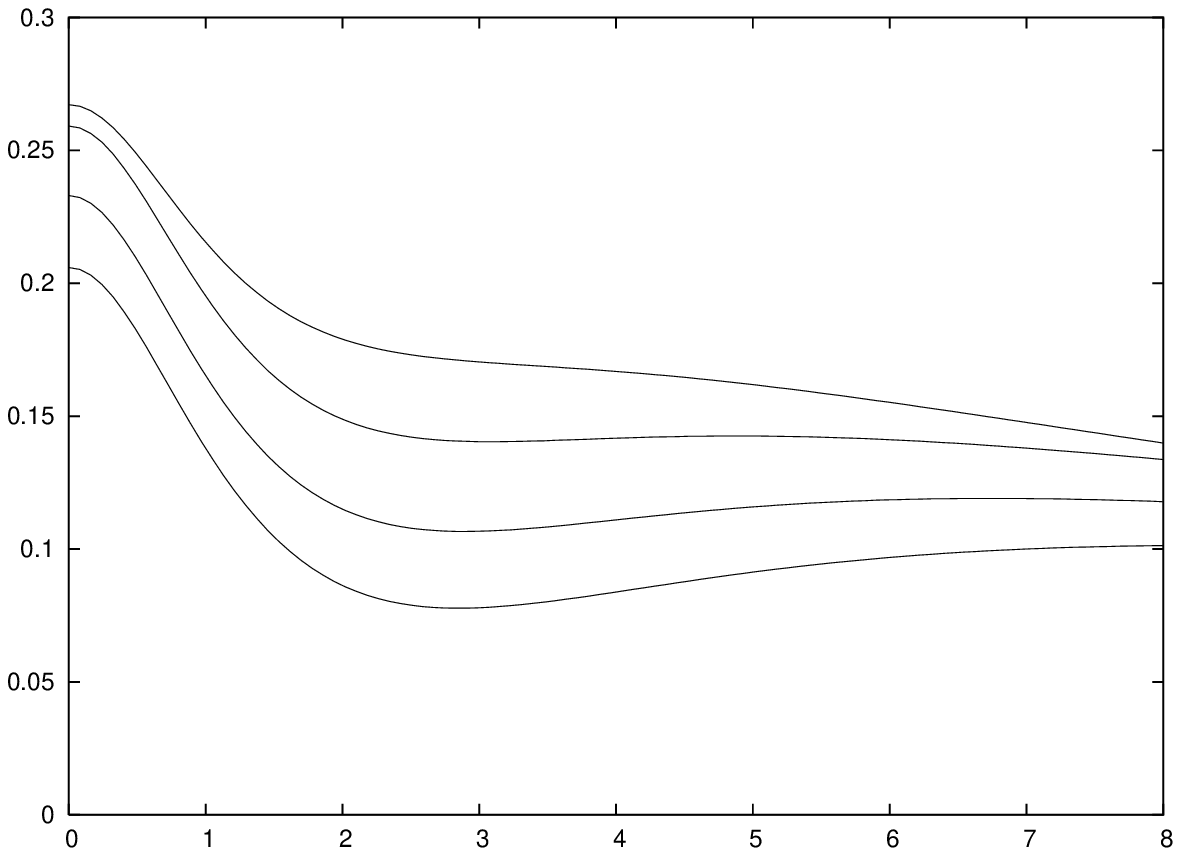} &
\epsfig{width=2.0in,file=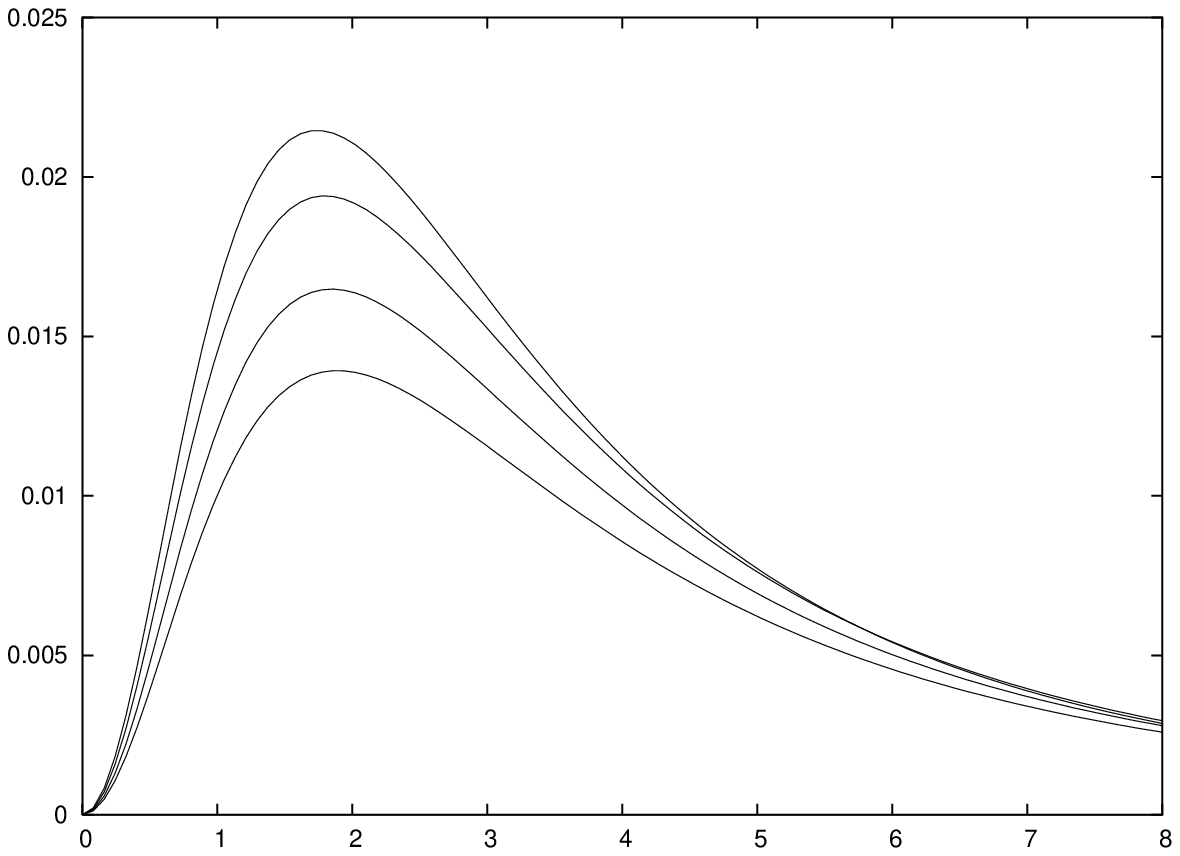}  \\
\tilde r & \tilde r & \tilde r \\
(d) & (e) & (f)  
\end{array}
$$	
\caption{ The electric charge density $\tilde {\mbox{\sl j} }_t$ 
$(a)$ for charged and magnetized Kerr-NUT disks  with  
$b_1=0.2$, $b_2=0.9$, $(b)$ for charged and magnetized Taub-NUT disks  with  $b_2=0.9$, and  $\alpha = 2.0$,  $c=1.0$
(axis $\tilde r$), $1.5$ (top curves),  $2.0$, $2.5$, and  $3.0$ (bottom curves), 
and $(c)$ for Kerr-Newman disks with $b_1=0.2$, and 
$\alpha = 2.0$,  $c=1.0$
(axis $\tilde r$), $2.0$ (top curve),  $1.5$, $2.5$, and  
$3.0$ (bottom curve), and the  azimuthal
current density $\mbox{\sl j}_\varphi$ $(d)$ for charged and magnetized Kerr-NUT disks  with  $b_1=0.2$, $b_2=0.9$, 
$(e)$ for charged and magnetized Taub-NUT disks  
with  $b_2=0.9$, and $(f)$ for Kerr-Newman disks with 
$b_1=0.2$, and $\alpha = 2.0$,  $c=1.0$
(axis $\tilde r$), $1.5$ (top curves),  $2.0$, $2.5$, and  
$3.0$ (bottom curves),
as functions of  $\tilde r$.} \label{fig:j0j1}
\end{figure*}

The electric charge density $ \tilde {\mbox{\sl j}}_t $ and  the  azimuthal
current density $\mbox{\sl j}_\varphi$, measured  in the coordinates frame, 
are  represented in Fig.  \ref{fig:j0j1}. We observe
that the electric charge density has a  similar behavior to the energy density  which is consistent with the
fact that the mass  is more concentrated in the disks center. We also computed 
this functions  for other values of the parameters and, in  all the cases, we
found  the same behavior. 


Now we will interpret the matter in the disks as two  
counterrrotating  streams of charged  dust particles (CRM).  First we consider the two 
streams of particles  moving on  electrogeodesics  
for charged and magnetized Kerr-NUT disks with 
$b_1=0.2$, $b_2=0.9$, and Kerr-Newman disks with $b_1=2$, 
for $\alpha = 1.2$, and $c=1.0$, $1.1$,  $1.3$, $1.4$,
as functions of $\tilde r$. For charged and magnetized Taub-NUT fields we found  a similar behavior  to the charged and magnetized Kerr-NUT fields. 

\begin{figure*}
$$
\begin{array}{cccc}
  \omega _+& \omega _-  & \omega_+  &\omega_-   \\ 
\epsfig{width=1.5in,file=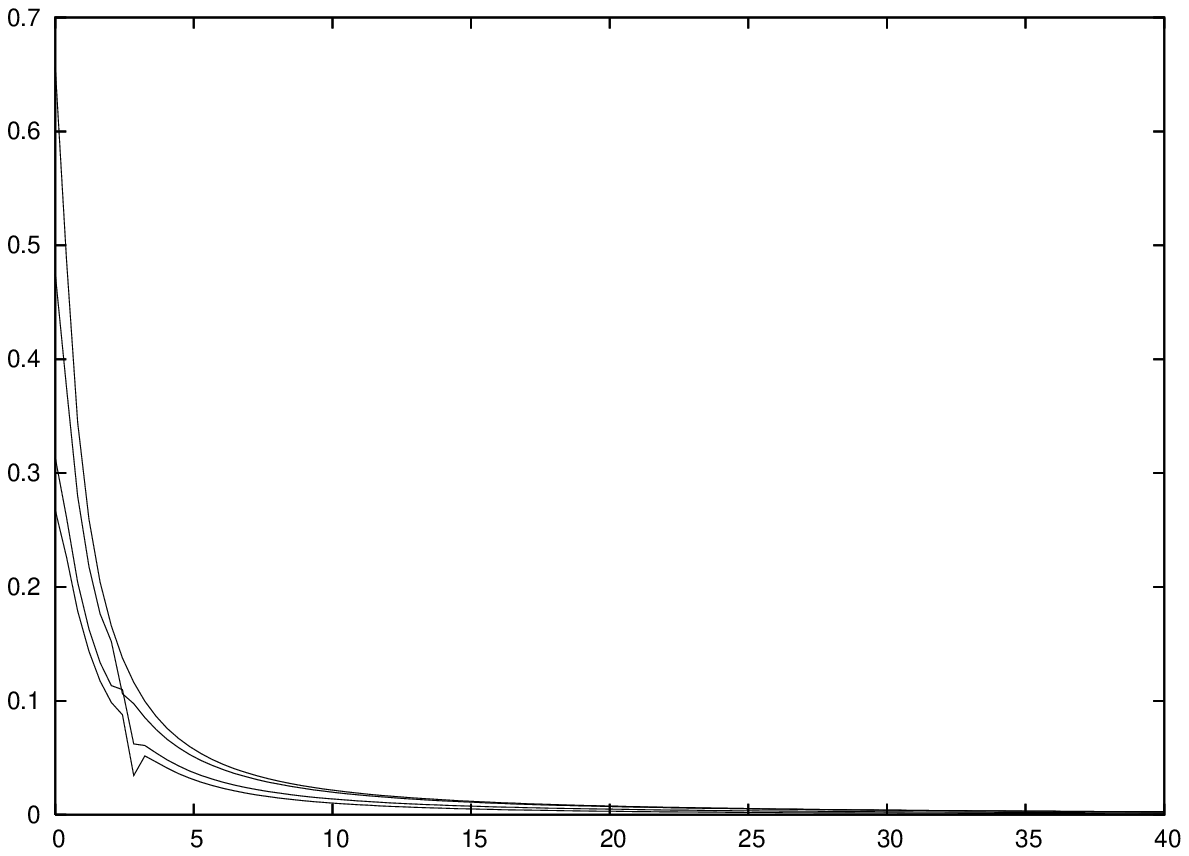} &
\epsfig{width=1.5in,file=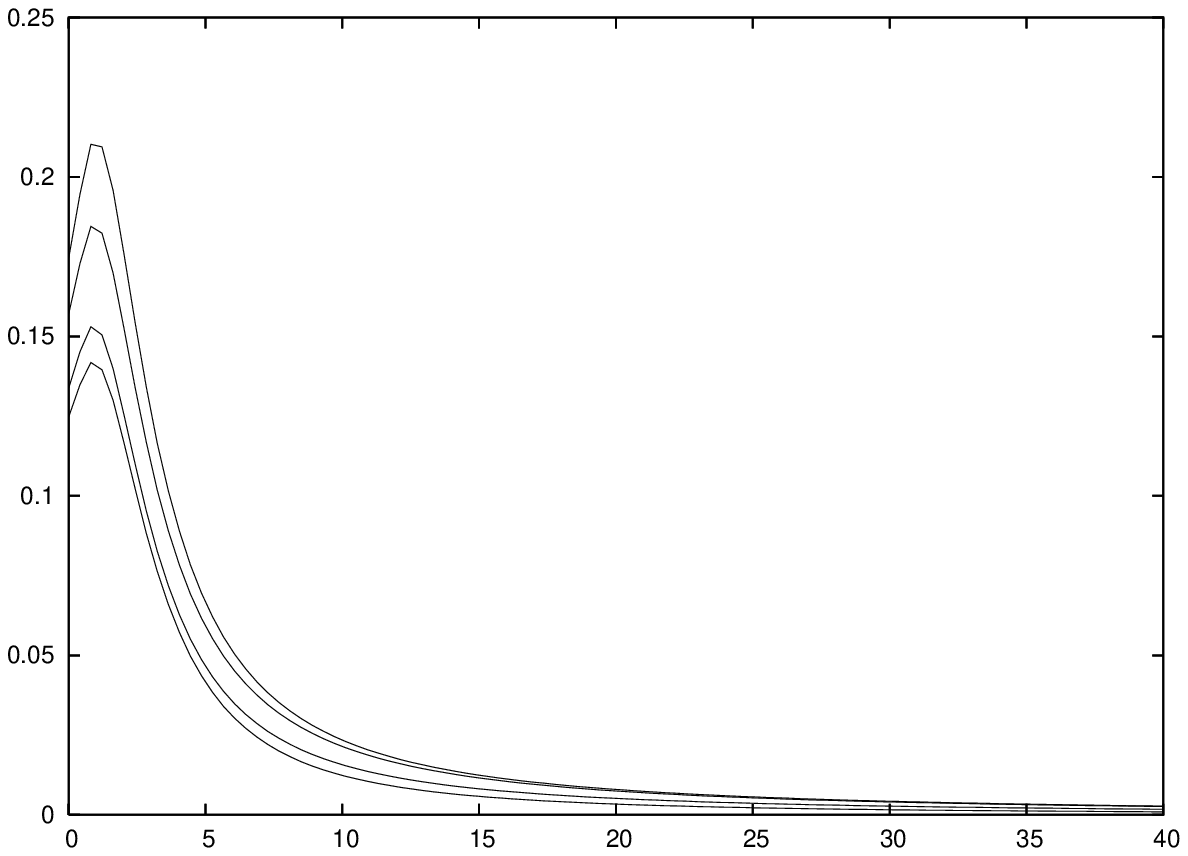} &
\epsfig{width=1.5in,file=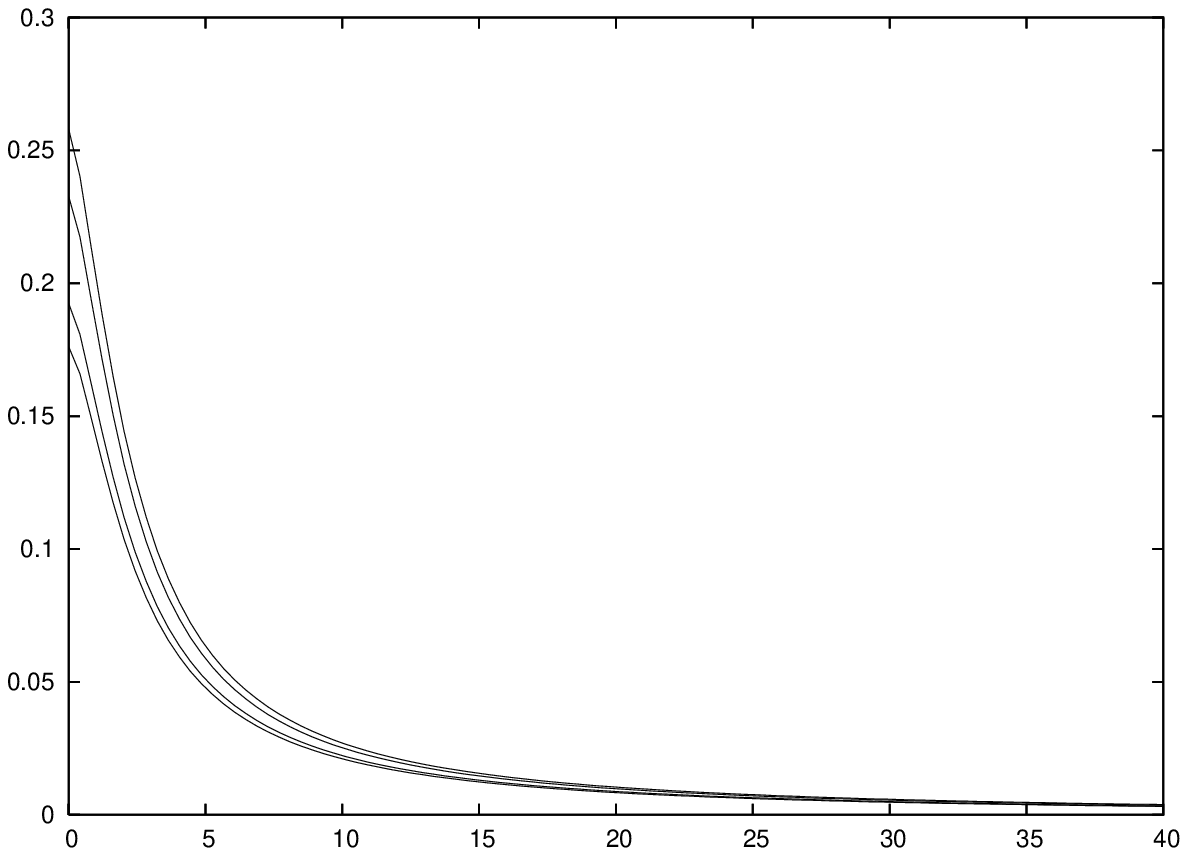}  &
\epsfig{width=1.5in,file=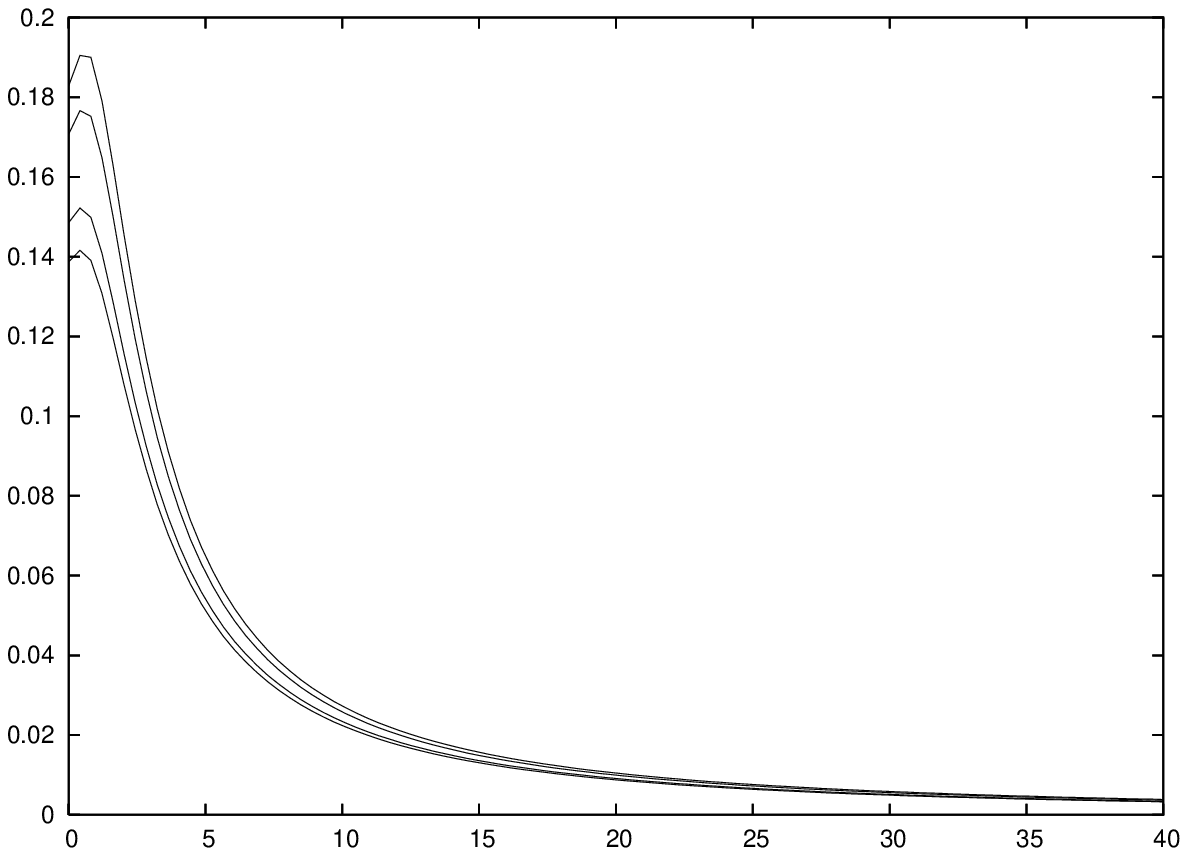}  \\
\tilde r & \tilde r & \tilde r & \tilde r \\
(a) & (b) & (c) & (d)   \\ 
\tilde \epsilon _+ & \tilde \epsilon _- & 
\tilde \epsilon _+ & \tilde \epsilon _-  \\
\epsfig{width=1.5in,file=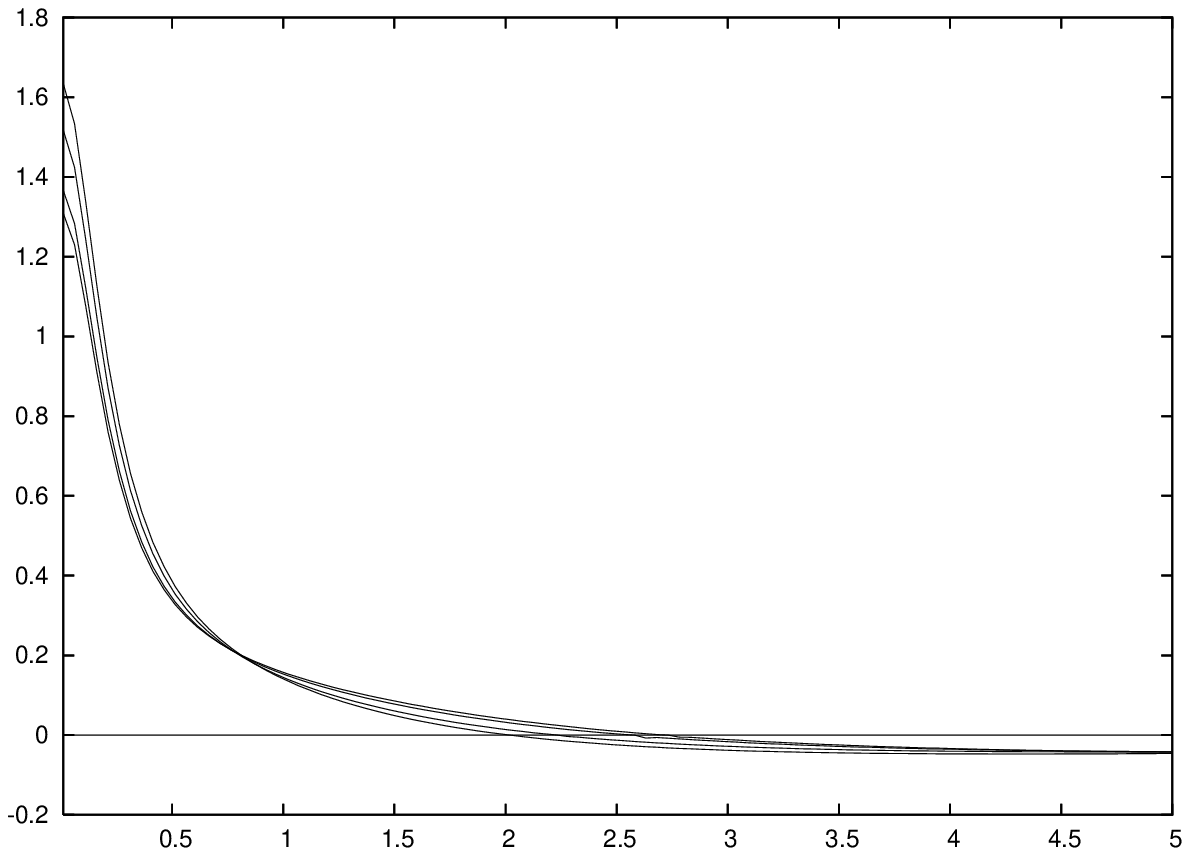} &
\epsfig{width=1.5in,file=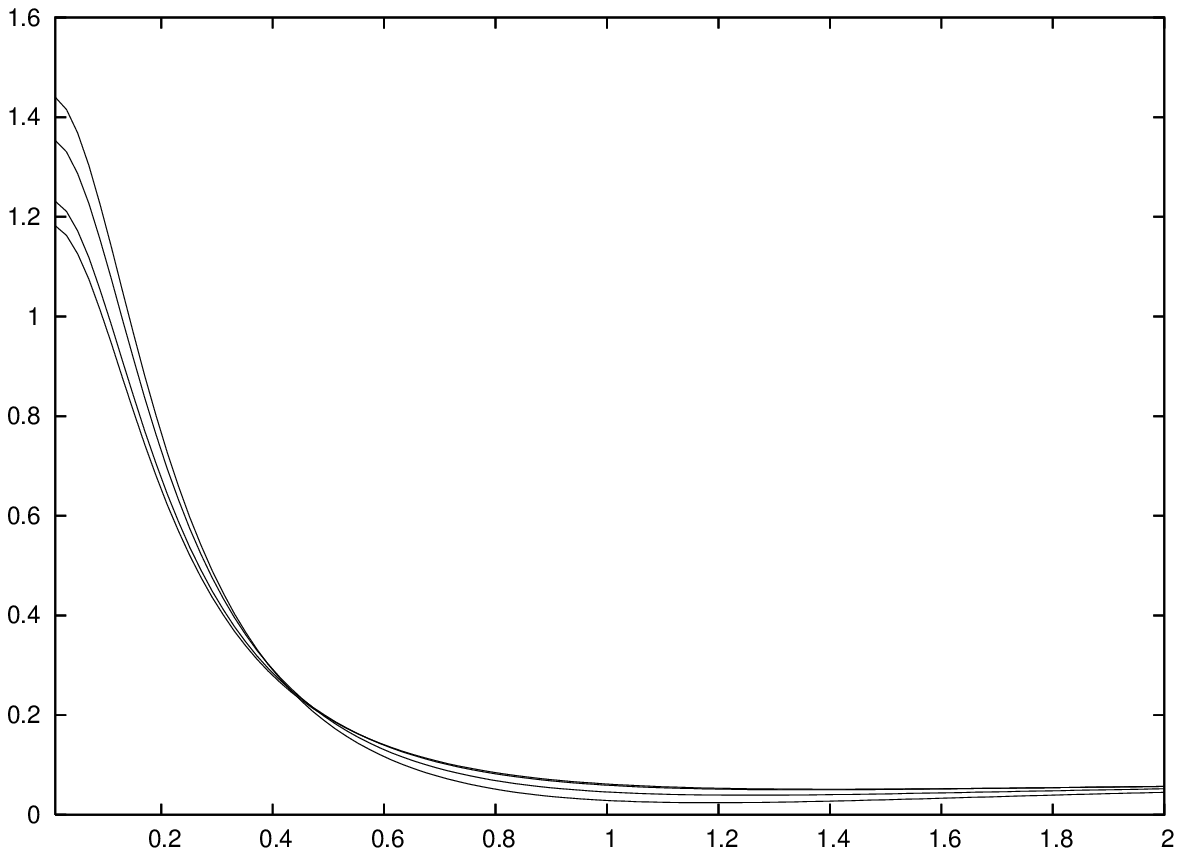} &
\epsfig{width=1.5in,file=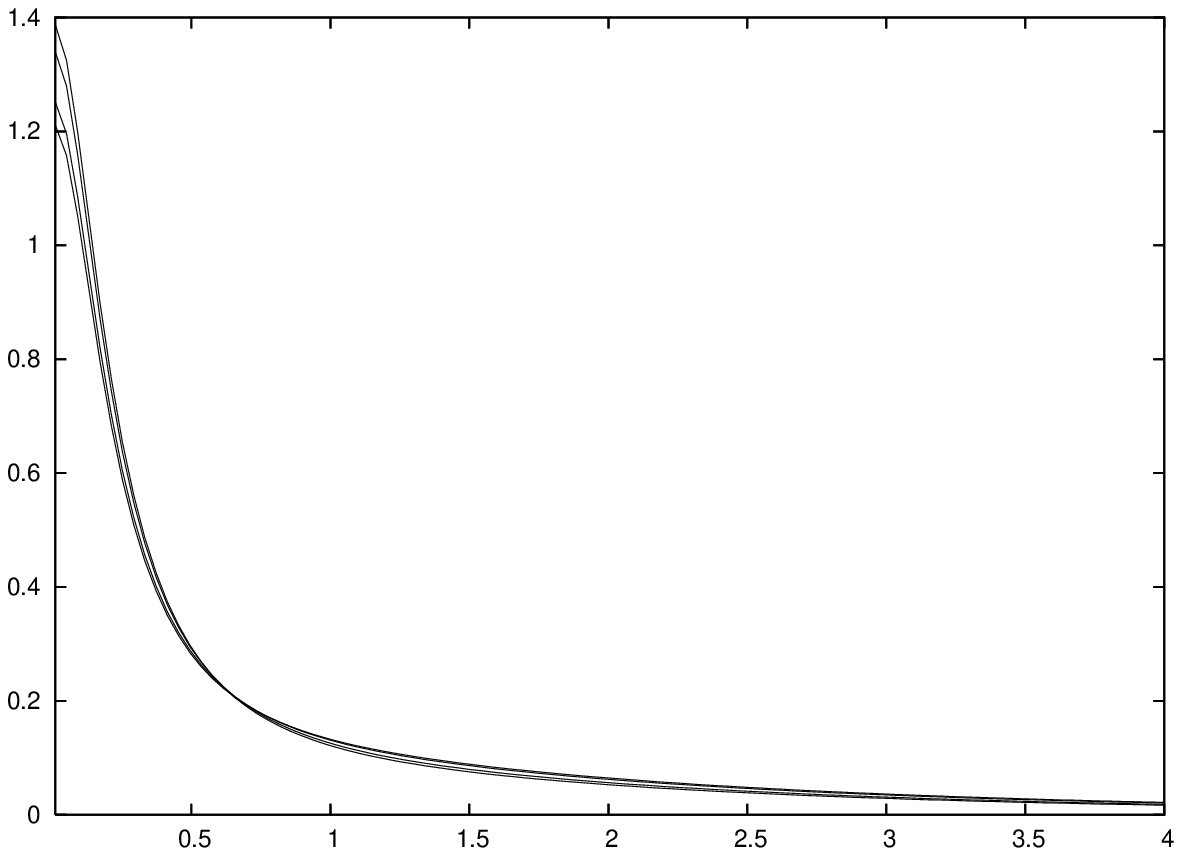} &
\epsfig{width=1.5in,file=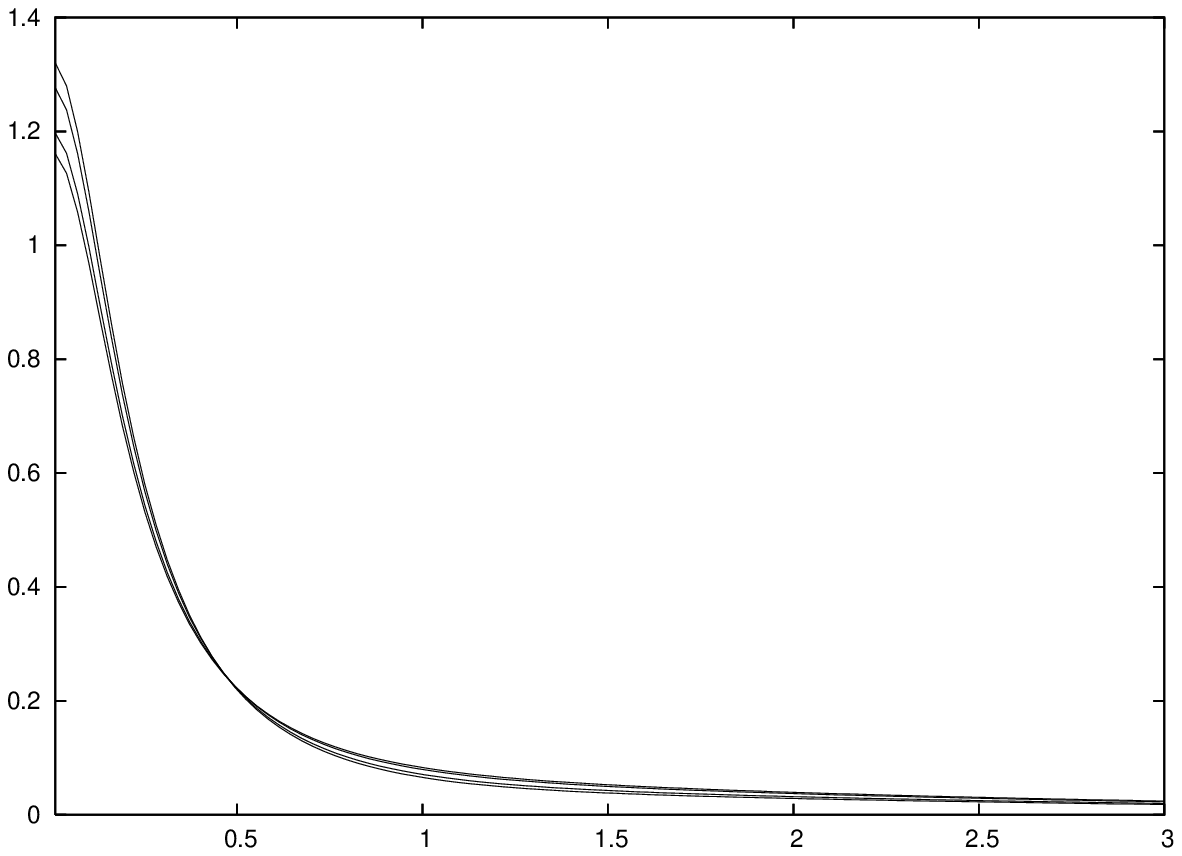}  \\
\tilde r & \tilde r & \tilde r & \tilde r \\
(e) & (f) & (g) & (h)   \\ 
-\tilde \sigma _+ & \tilde \sigma _- &
-\tilde \sigma _+ & \tilde \sigma _-  \\
\epsfig{width=1.5in,file=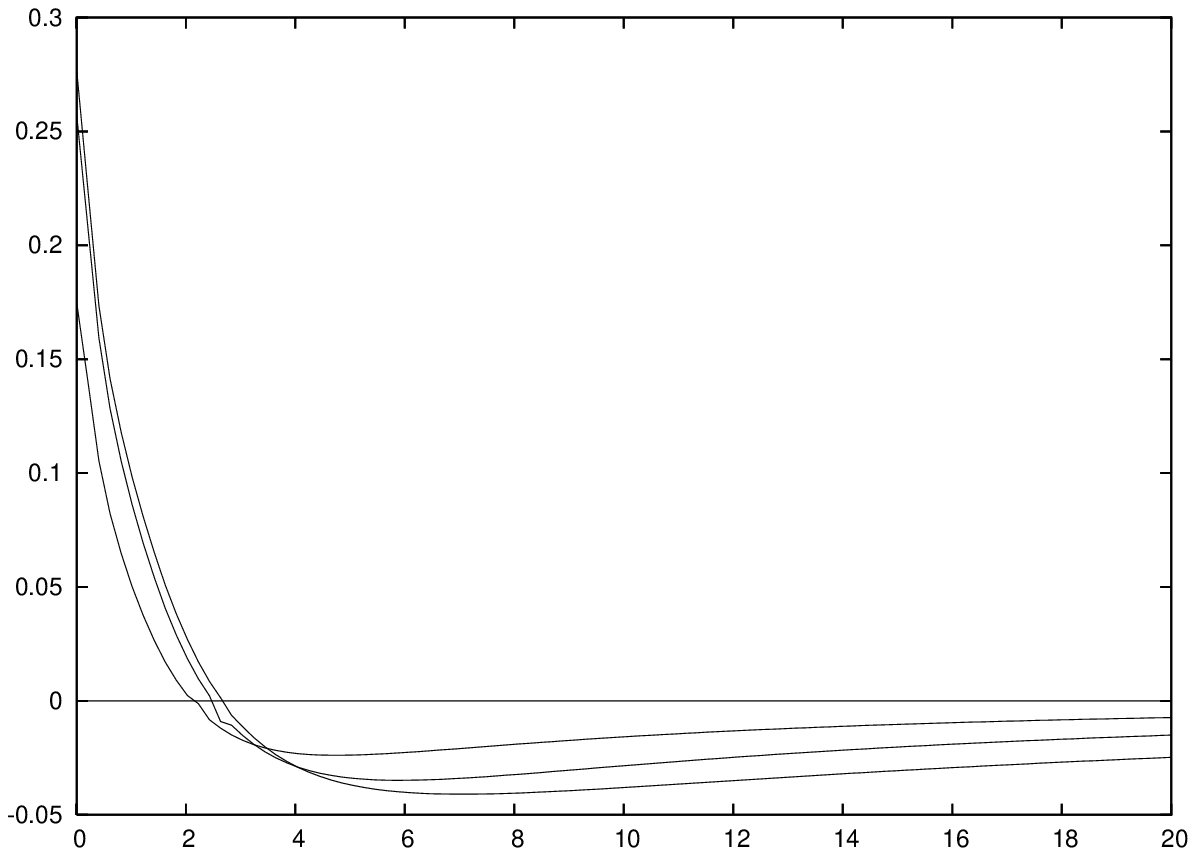} &
\epsfig{width=1.5in,file=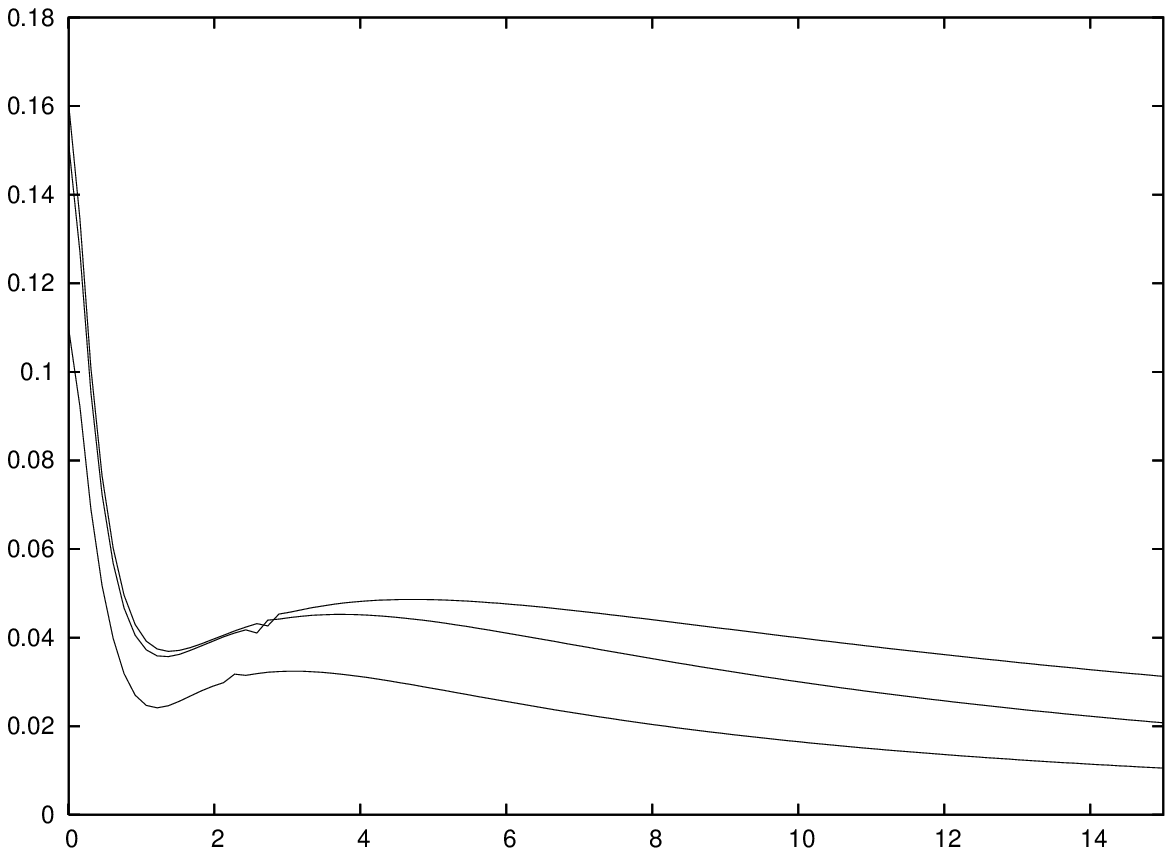} &
\epsfig{width=1.5in,file=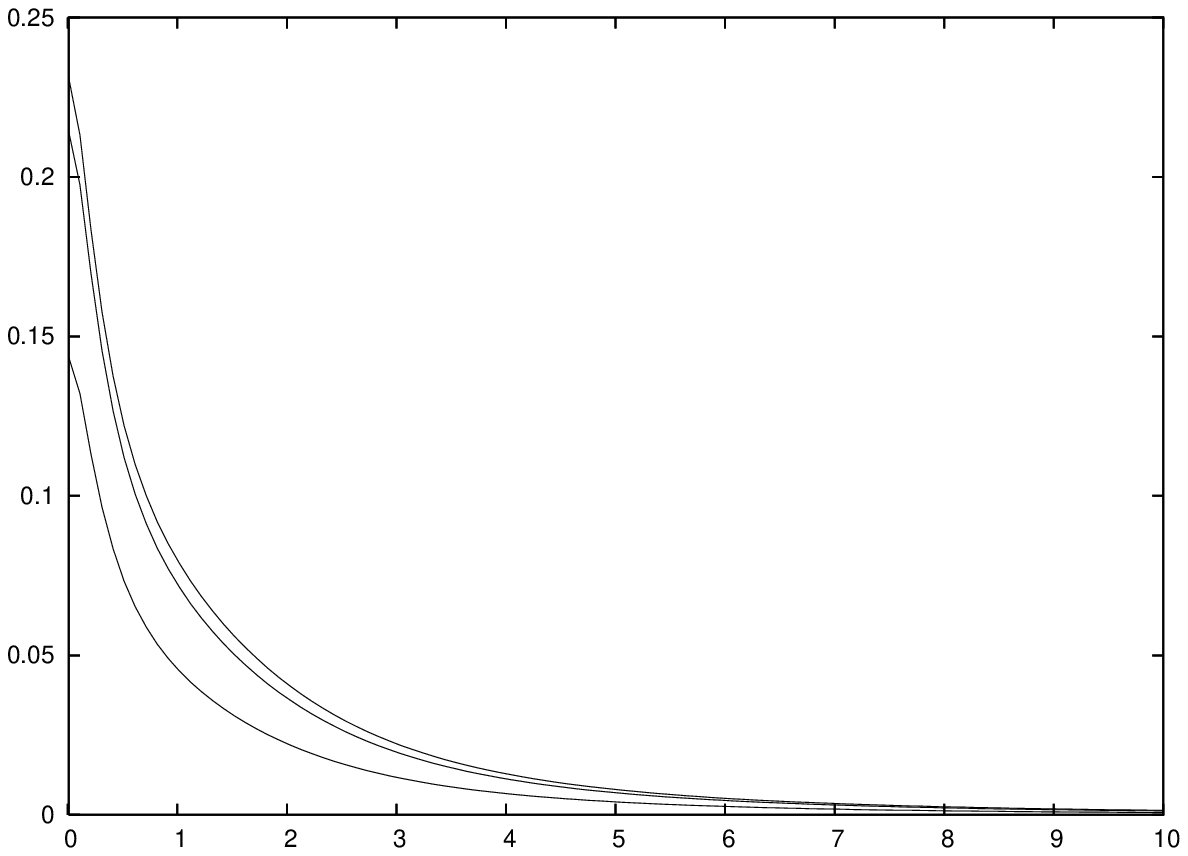} &
\epsfig{width=1.5in,file=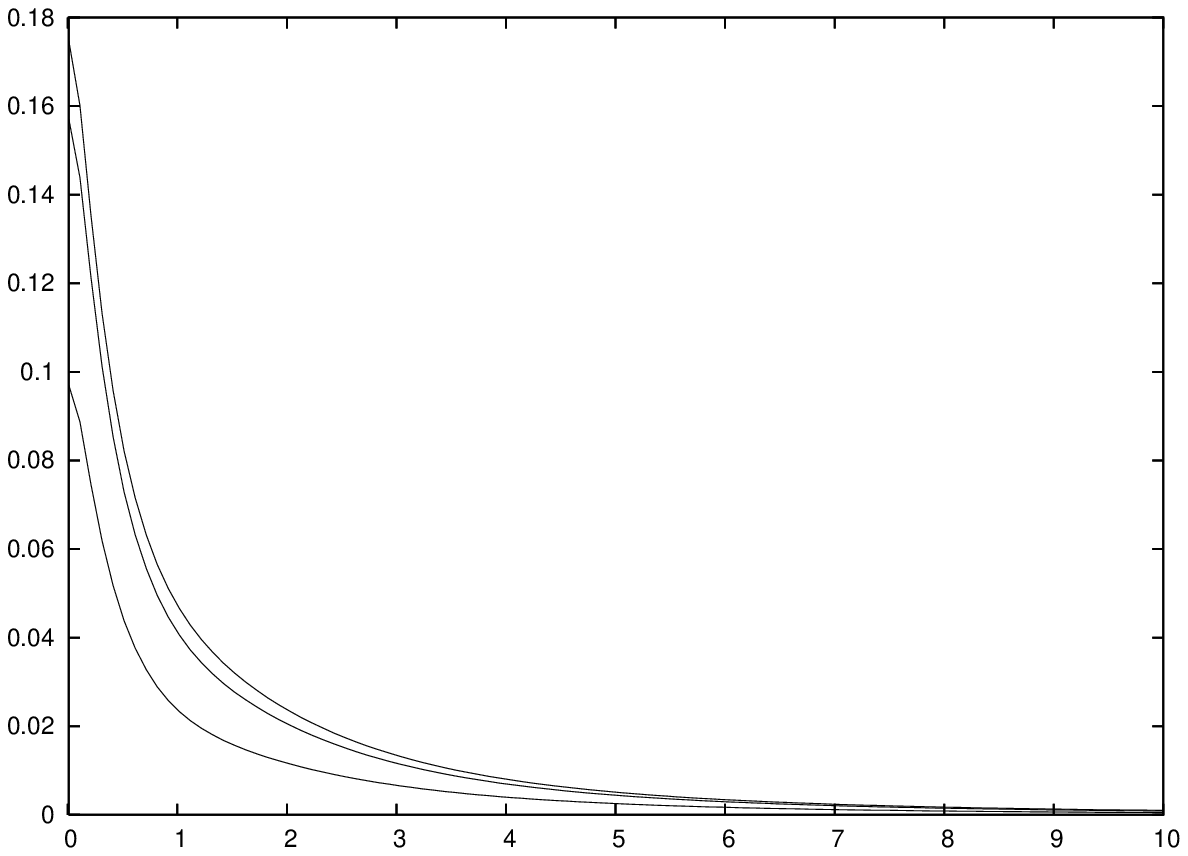} \\
\tilde r & \tilde r & \tilde r & \tilde r \\
(i) & (j) & (k) & (l)   \\ 
h^2_+ & h^2_- & h^2_+ & h^2_-  \\
\epsfig{width=1.5in,file=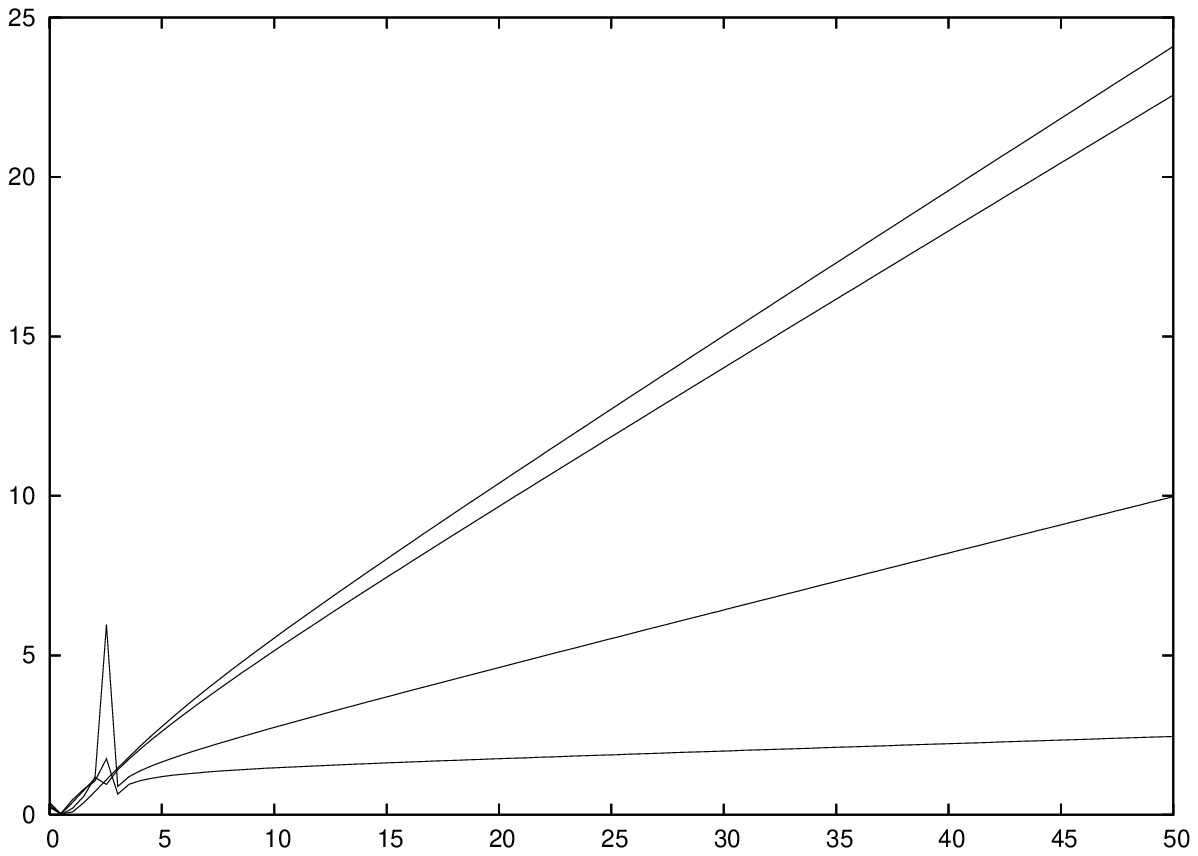} &
\epsfig{width=1.5in,file=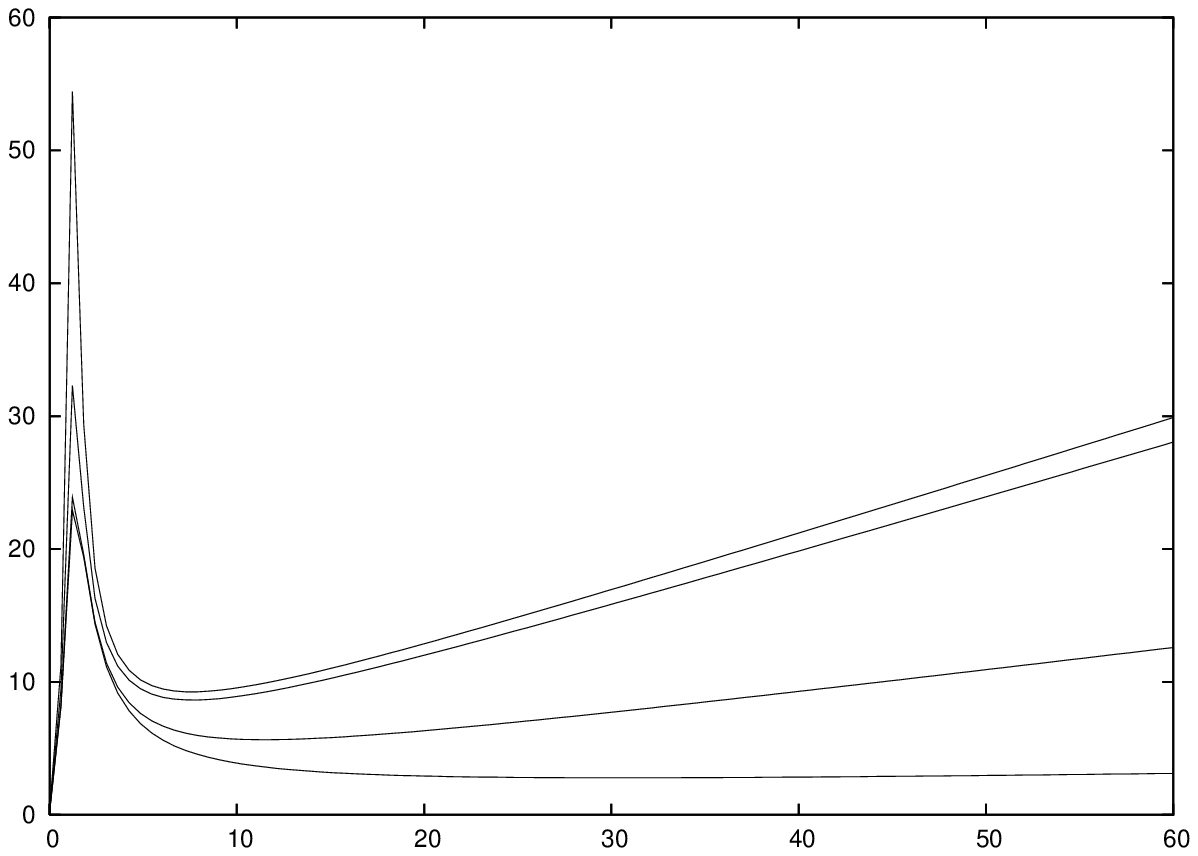} &
\epsfig{width=1.5in,file=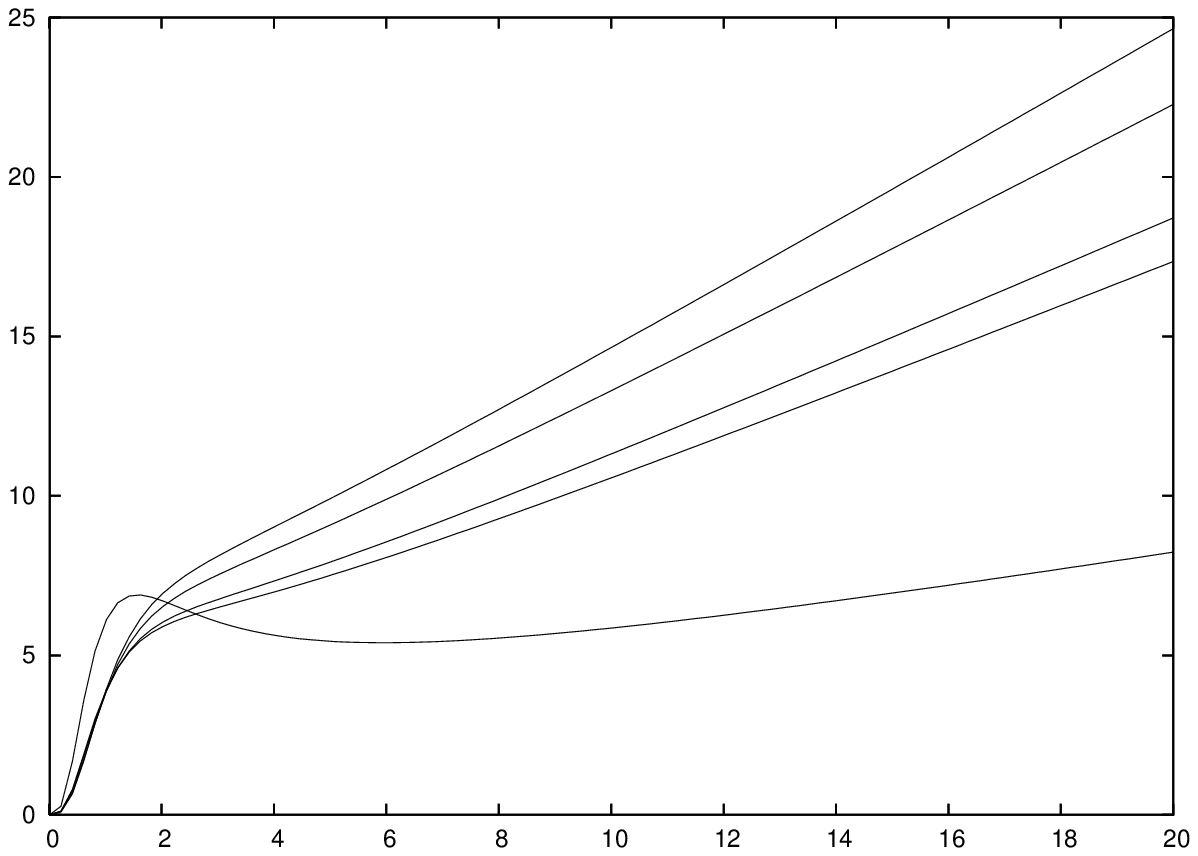} &
\epsfig{width=1.5in,file=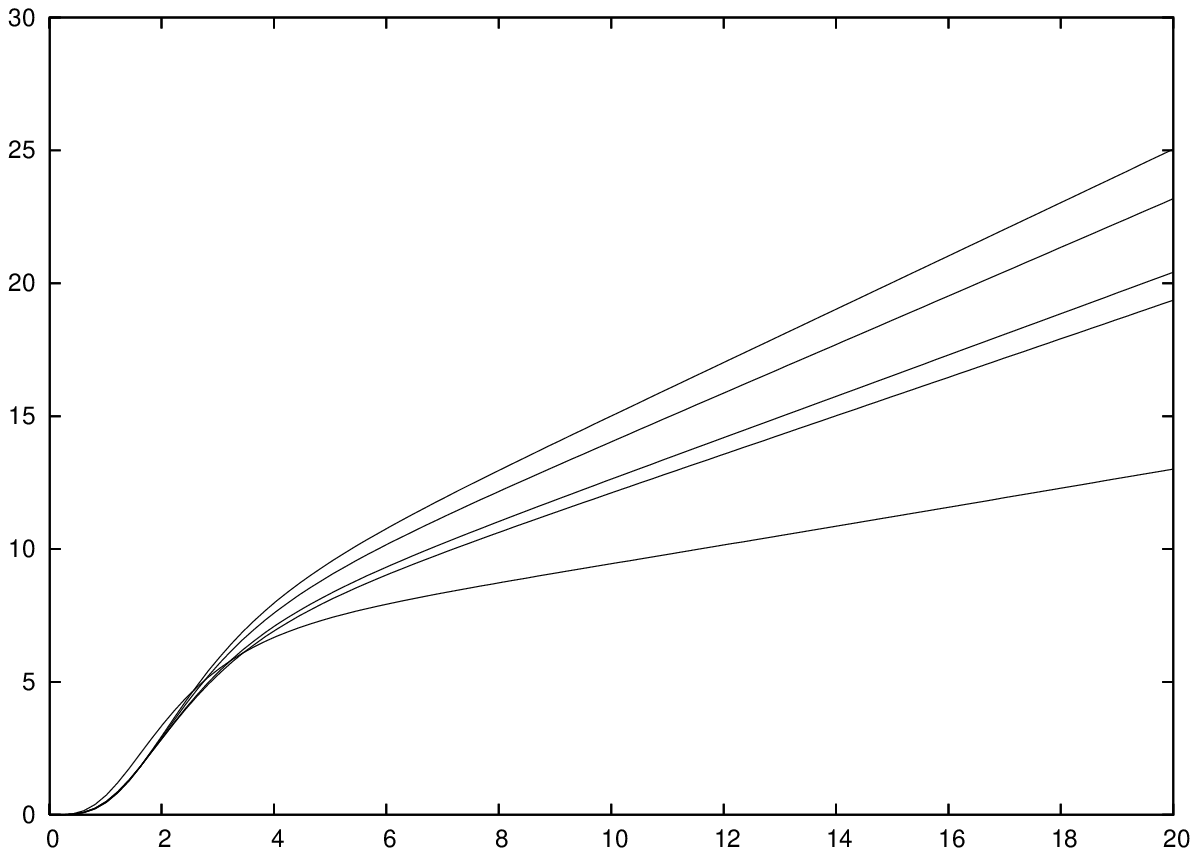} \\
\tilde r & \tilde r & \tilde r & \tilde r \\
(m) & (n) & (o) & (p)         
\end{array}
$$
\caption{We plot for electrogeodesic charged and magnetized
Kerr-NUT fields with $\alpha=1.2$, $b_1=0.2$, $b_2=0.9$, and  
electrogeodesic Kerr-Newman fields  with  $\alpha = 1.2$, 
$b_1=0.2$, 
the angular velocities $\omega _\pm$ $(a)$, $(b)$
for charged and magnetized Kerr-NUT disks and  $(c)$, $(d)$
for Kerr-Newman disks, the energy densities $\tilde 
\epsilon _\pm$  $(e)$, $(f)$
for charged and magnetized Kerr-NUT disks and $(g)$, $(h)$
for Kerr-Newman disks, with   $c=1.0$ (top curves), $1.1$,  
$1.3$, $1.4$ (bottom curves), the electric charge 
densities $\tilde \sigma_ \pm$  $(i)$, $(j)$
for charged and magnetized Kerr-NUT disks and $(k)$, $(l)$
for Kerr-Newman disks, with   $c=1.0$ (axis $\tilde r$), $1.1$ 
(bottom curves),  $1.3$, $1.4$ (top curves), the specific angular
momenta $h^2_\pm$ $(m)$, $(n)$
for charged and magnetized Kerr-NUT disks and  $(o)$, $(p)$
for Kerr-Newman disks, with   $c=1.0$ (top curves), 
$1.1$,  $1.3$, $1.4$, $3.0$ (botton curves). }
\label{fig:electro}
\end{figure*}

In Figs. \ref{fig:electro}$(a)$  
- \ref{fig:electro}$(d)$  we plot the angular velocities curves
$\omega _\pm$.
We see that these  velocities  are real and for charged and
magnetized Kerr-NUT field $\omega _+$  presents a nonsmooth 
behavior at  certain value of $\tilde r$.  
In Figs. \ref{fig:electro}$(e)$ - \ref{fig:electro}$(h)$   
we have  plotted the energy densities
$\tilde \epsilon_\pm$. We see that 
for charged and magnetized Kerr-NUT disks $\tilde \epsilon_-$ 
is everywhere positive whereas $\tilde
\epsilon_+$ becomes negative after certain value of
$\tilde r$ in violation of the weak energy condition.
Therefore for charged and magnetized Kerr-NUT fields one finds that only the central region of the these disks presents
a physically reasonable behavior. Thus these disks only model the inner portions of galaxies or acretion disks. 
However, since the surface energy density decreases rapidly 
one can  to define a cut off radius at the point where the
energy density changes of sign and, in principle, we  consider these disks as finite so that we can use them to model  the whole galaxy or  acretion disk. By contrast
for Kerr-Newman disks
the counterrotating  energy densities are always positive 
quantities. 

In Figs. \ref{fig:electro}$(i)$ - 
\ref{fig:electro}$(l)$ we plotted the electric
charge densities $\tilde \sigma _\pm$.
We also find that these quantities have  a similar behavior
to $\tilde \epsilon _\pm $.
In Figs. \ref{fig:electro}$(m)$ - \ref{fig:electro}$(p)$   
we have  drawn the  specific angular
momenta  $h^2_\pm$  for the same values  of the 
parameters. We see that for charged and magnetized Kerr-NUT
disks  there is a strong change in the
slope  at certain values of  $\tilde r$, which means that 
there is a strong
instability there. We also find regions with negative slope 
where the CRM is also unstable.  
However, these disks models are stable in the central region
and  away of the center of the disks.  
Meanwhile for Kerr-Newman disks
we find  that the
presence of  electromagnetic field can also make unstable 
these disks  against
radial  perturbations. Thus the CRM cannot apply for $c=3.0$ 
(bottom curves). We also computed this functions  for other values of $b_1$ and $b_2$ and, in  all the cases, we
found  the same behavior.

\begin{figure*}
$$
\begin{array}{cc}
 v^2 & \tilde \epsilon _\pm    \\ 
\epsfig{width=2.5in,file=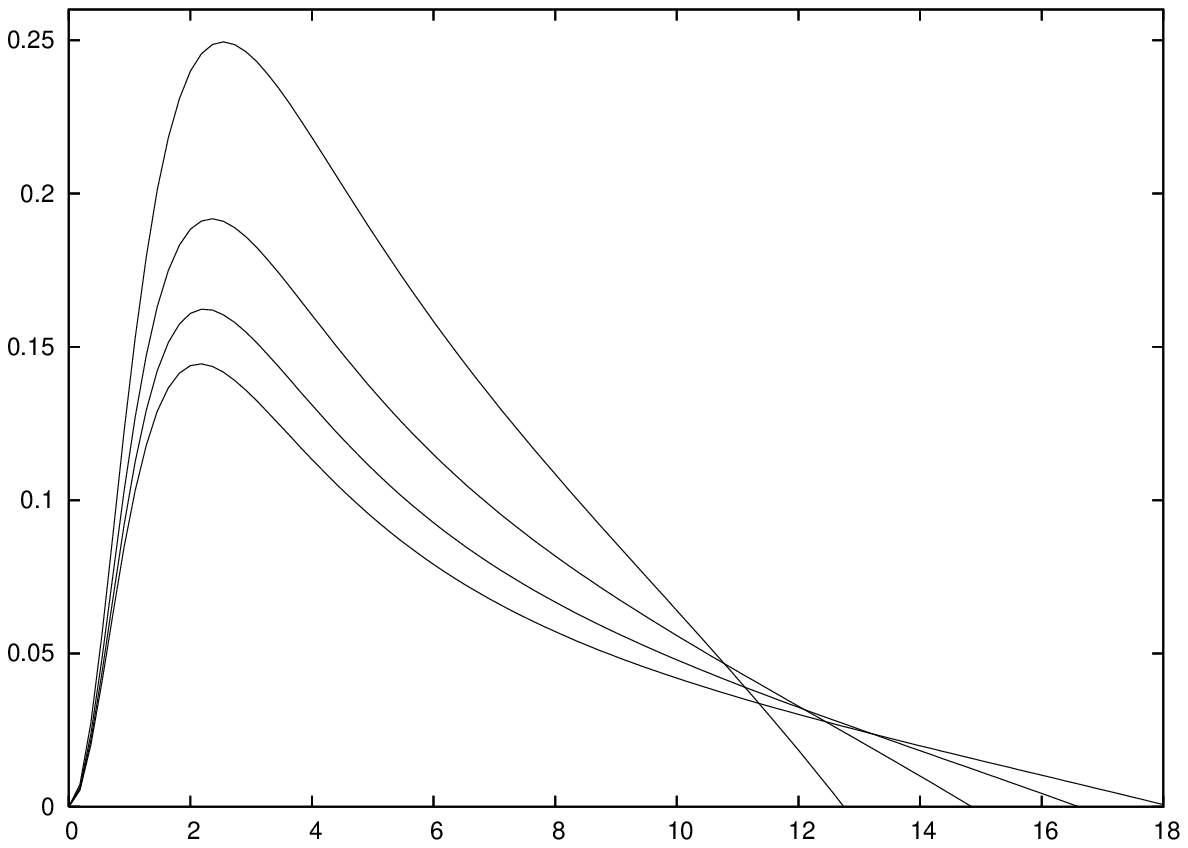}     &
\epsfig{width=2.5in,file=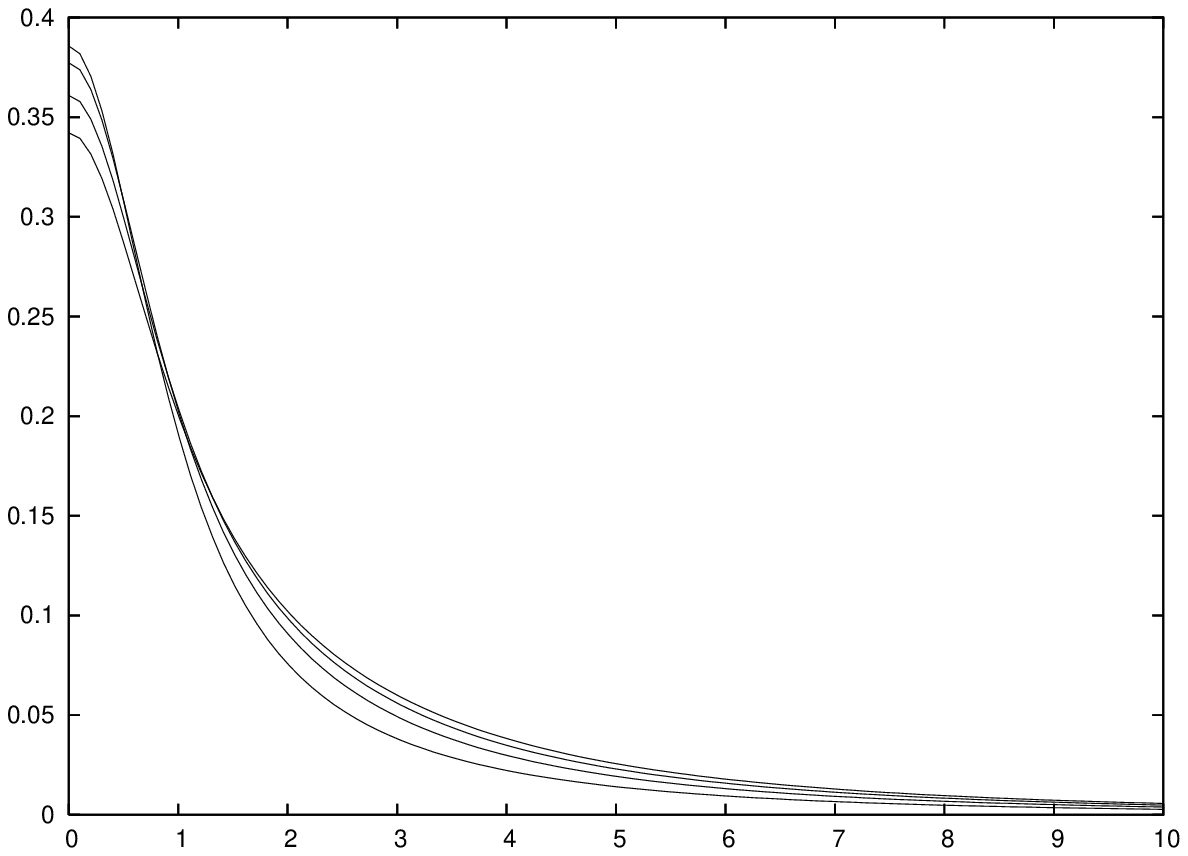}    \\
\tilde r  & \tilde r  \\
(a)  &    (b)  \\
-\tilde \sigma _+   &  -\tilde \sigma _-  \\
\epsfig{width=2.5in,file=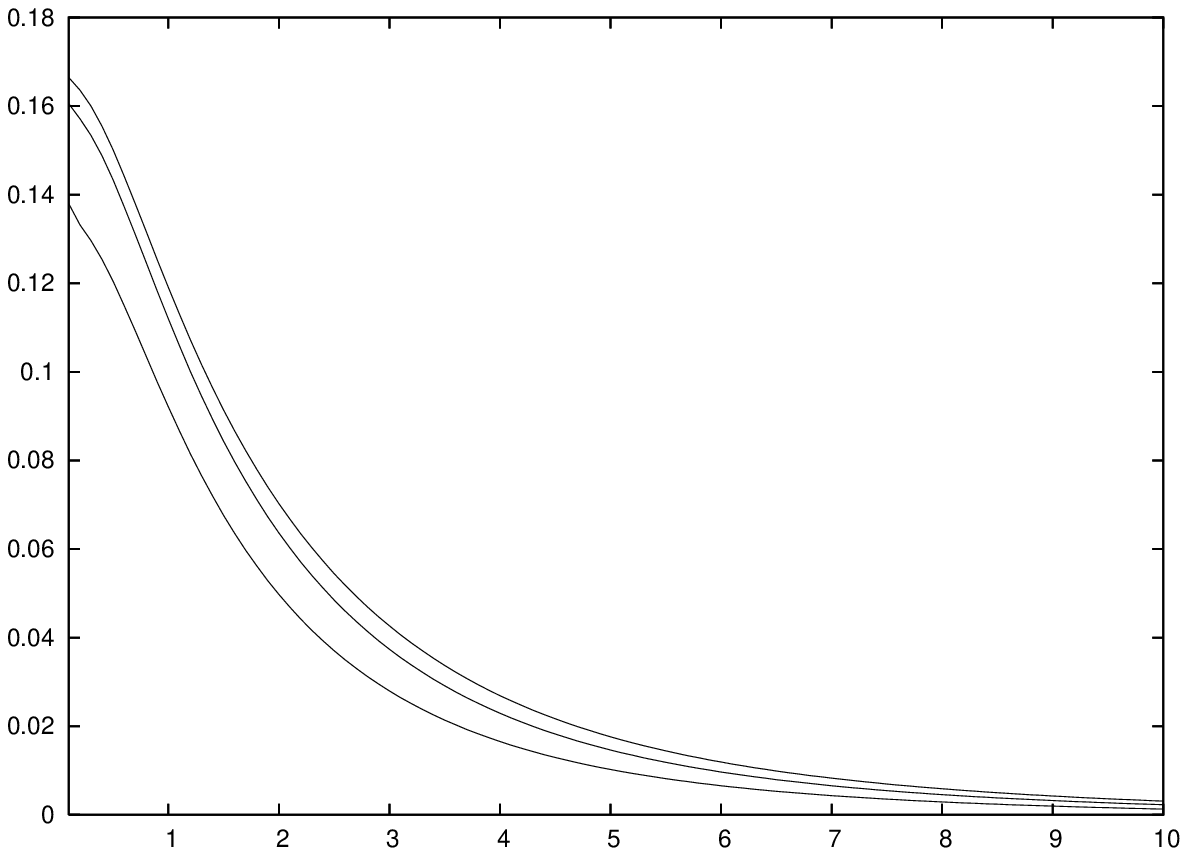}  &
\epsfig{width=2.5in,file=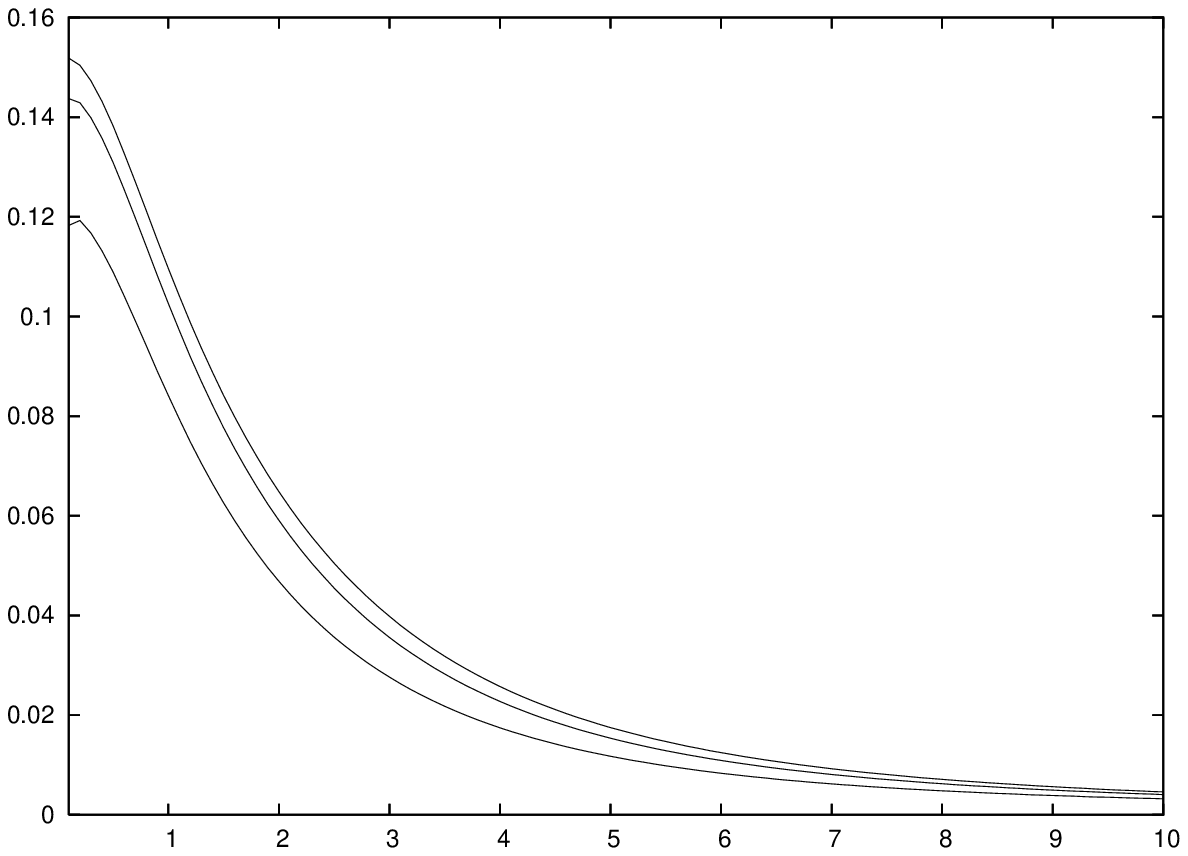} \\
\tilde r & \tilde r \\
(c)  & (d)   \\
 h^2_+   &     h^2_-     \\
\epsfig{width=2.5in,file=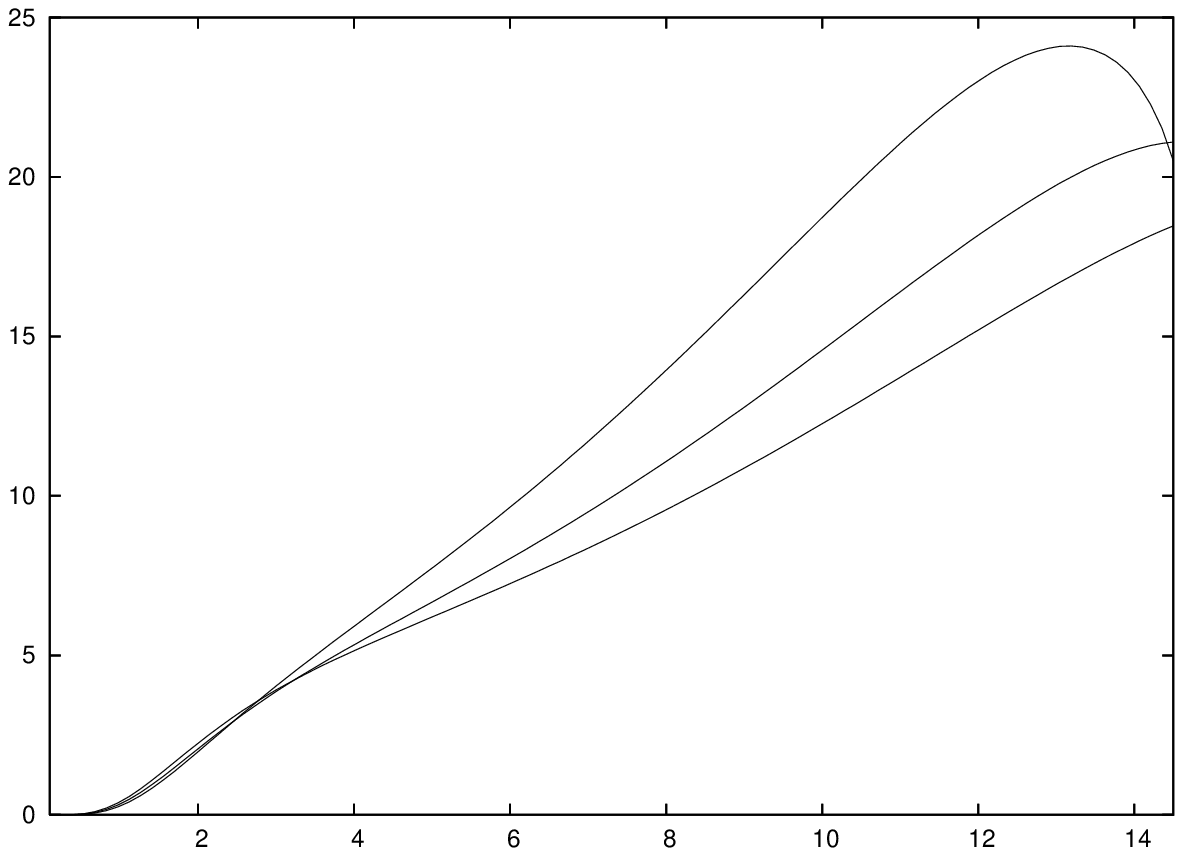} &
\epsfig{width=2.5in,file=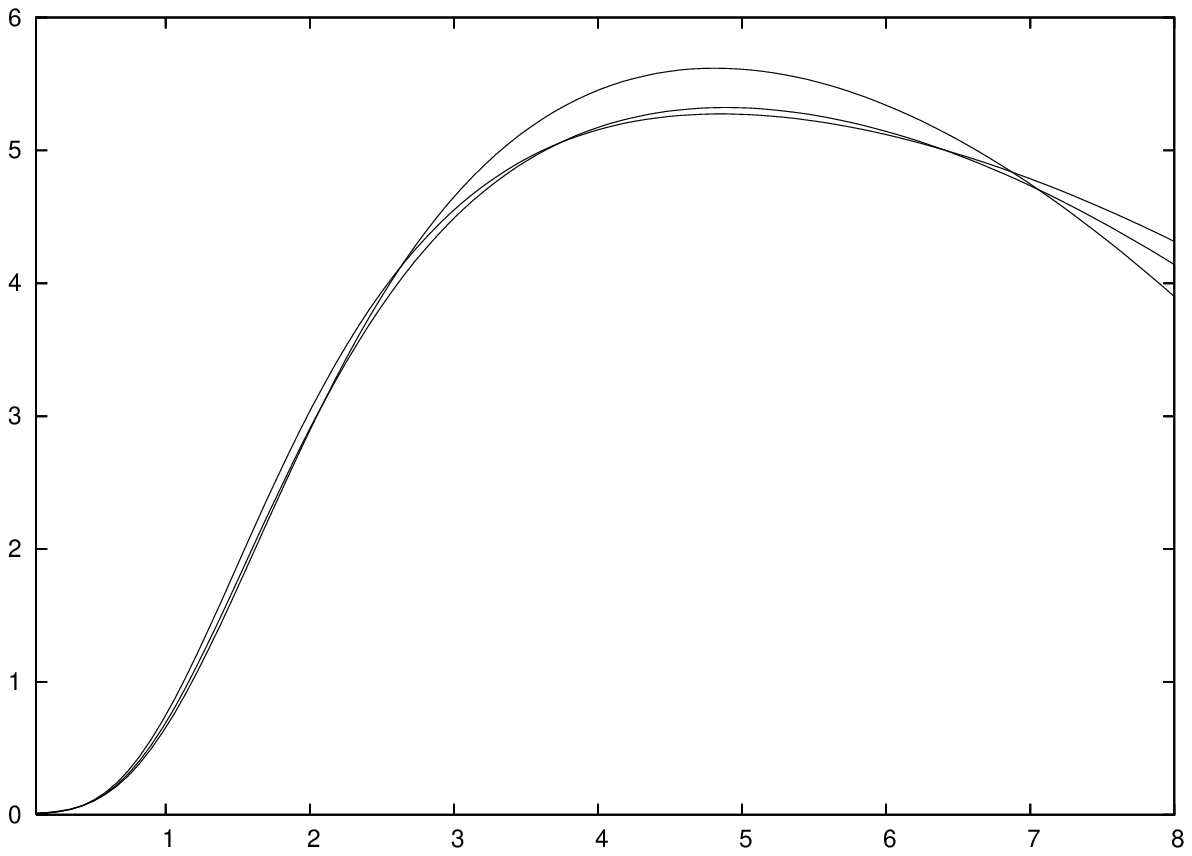}    \\
\tilde r & \tilde r  \\
(e)    &    (f)  
\end{array}
$$	
\caption{We plot  for non-electrogeodesic charged and
magnetized  Kerr-NUT field with $\alpha=2$, 
$b_1=b_2=0.1$, $(a)$ the tangential velocity 
$v^2$, $(b)$ the energy densities $\tilde \epsilon _\pm $
for $c=1.0$ (top curves), $1.5$, $2.0$, $2.5$  (bottom curves), 
$(c)$, $(d)$ the electric charge densities 
$\tilde  \sigma _\pm $  for $c=1.0$ (axis $\tilde r$), 
$1.5$ (bottom curves), $2.0$, $2.5$ (top curves), $(e)$ and
$(f)$ the angular momentum specific $h^2_\pm$ for
$c=1.0$ (top curves), $1.5$, $2.0$, $2.5$  (bottom curves),
as functions of $\tilde r$.}  
\label{fig:noelectro}
\end{figure*}

Second we consider the case when  the two fluids move with equal 
and opposite tangential velocities (non-electrogeodesic motion) 
for charged and magnetized Kerr-NUT disks with   $\alpha=2$, 
$b_1=b_2=0.1$ and  $c=1.0$, $1.5$, $2.0$, $2.5$. 
In Fig. \ref{fig:noelectro}$(a)$ we plot the  tangential velocity
curves of the counterrotating streams, $v^2$. We see that the 
quantity $v^2$ is also always
less than the light velocity, but after of certain value of
$r$ becomes negative. Therefore these disks are also well behaved 
only  in the central region of the disk.
In Figs. \ref{fig:noelectro}$(b)$ - \ref{fig:noelectro}$(d)$  we
have  plotted the energy densities $\tilde \epsilon _\pm$ and
the electric charge densities $\sigma _\pm $.
We see that this
quantities have  a similar behavior to the previous cases. 
In Figs. \ref{fig:noelectro}$(e)$ and 
\ref{fig:noelectro}$(f)$ we have  drawn the  specific angular 
momena  $h^2_+$ and 
$h^2_-$ for the same values  of the parameters.
We find  that these disks models are stable only in the  
central region  of the disks. 
In the case of charged and magnetized Taub-NUT
disks  we also find that the physical quantities 
have a similar  behavior to the
charged and magnetized Kerr-NUT disks, and in  the case
of Kerr-Newman disks  to the previous ones. 

Finally, the Figs. \ref{fig:disenprq} - \ref{fig:noelectro} show that the two fluids
are continuous in $r$ which implies to have  two particles in counterrotating
movement in the same point in spacetime. So this model could be possible when
the distance between streams (or between the counterrotating particles) were
very small in comparing with the length  $r$  so that we can consider, in
principle, the fluids continuous  like is the case of  counterrotating gas
disks present  in disk galaxies.


\section{Discussion}

We presented a detailed analysis of the energy-momentum tensor and the surface
current density for electrovacuum stationary axially symmetric relativistic
thin disks without radial stress and when there is  heat
flow. The surface
energy-momentum tensor and the surface current density were expressed in terms
of the comoving tetrad and explicit expressions were obtained for the
kinematical and dynamical variables that characterize the disks. That is, we
obtained  expressions for the velocity vector of the disks, as well for the
energy density, azimuthal pressure, electric charge density and azimuthal current density.

We also presented in this paper the stationary generalization of the 
counterrotating model (CRM) for electrovacuum thin disks that had been
previously only analyzed for the  static case in \cite{GG1,GG2}.   We considered both counter rotation with equal and opposite
velocities and counter rotation along electrogeodesics.  
A general constraint over the counterrotating tangential velocities was
obtained, needed to cast the surface energy-momentum tensor of the disk in such
a way that can be interpreted as the superposition of two counterrotating
dust  fluids. The constraint obtained is the generalization of the obtained
for the vacuum case in \cite{GL2}, for disks without radial pressure and with heat
flow, where we only consider counterrotating fluids circulating along
geodesics. We also found that, in general, there is not possible to take the
two counterrotating tangential velocities as equal and opposite neither take
the two counterrotating fluids as circulating along geodesics.

A family  of models of  counterrotating and rotating   relativistic thin disks  based on a charged and magnetized
Kerr-NUT metric are constructed 
where we obtain some disks with a CRM well behaved. 
This solution has as
limiting cases a charged and magnetized Taub-NUT solution 
and the well known  Kerr-Newman solutions. 
For charged and magnetized 
Kerr-NUT and Taub-NUT fields we find that 
these disks always present heat flow, 
whereas for Kerr-Newman fields  
we shows that these disks can
be interpreted,  for all the values of parameters, as a matter
distribution with currents and purely azimuthal pressure and 
without heat flow.
Finally, the inclusion of radial pressure to these 
models is being considered.








\newpage





\newpage


\newpage

\begin{thebibliography}{9999}

\bibitem{BS} W. A. Bonnor and A. Sackfield, Commun. Math.   Phys. {\bf 8}, 338
(1968).

\bibitem{MM1} T. Morgan and L. Morgan, Phys. Rev.  {\bf 183},  1097 (1969).

\bibitem{MM2} L. Morgan and T. Morgan, Phys. Rev. D  {\bf 2},  2756 (1970).

\bibitem{GL1} G. A. Gonz\'alez and P. S. Letelier, Class.  Quantum Grav. {\bf
16}, 479 (1999).

\bibitem{LP} D. Lynden-Bell and S. Pineault, Mon. Not. R.  Astron. Soc. {\bf
185}, 679 (1978). \label{bib:LP} 

\bibitem{CHGS} A. Chamorro, R. Gregory, and J. M. Stewart, Proc. R. Soc. London {\bf A413}, 251 (1987).

\bibitem{LO} P.S. Letelier and S. R. Oliveira, J. Math.  Phys.  {\bf 28}, 165 (1987).

\bibitem{LEM} J. P. S. Lemos, Class. Quantum Grav. {\bf 6}, 1219 (1989).

\bibitem{LL1} J. P. S. Lemos and P. S. Letelier, Class.  Quantum Grav. {\bf
10}, L75 (1993).

\bibitem{BLK} J. Bi\u{c}\'{a}k, D. Lynden-Bell, and J.  Katz,  Phys. Rev. D {\bf 47}, 4334 (1993).

\bibitem{BLP} J. Bi\u{c}\'{a}k, D. Lynden-Bell, and C.  Pichon, Mon. Not. R. Astron. Soc. {\bf 265}, 126 (1993).

\bibitem{GE} G.A. Gonz\'alez and O. A. Espitia,  Phys. Rev. D  {\bf 68}, 104028 (2003). \label{bib:GE}

\bibitem{BL} J. Bi\u{c}\'ak and T. Ledvinka, Phys. Rev.  Lett. {\bf 71}, 1669  (1993).

\bibitem{GL2} G. A. Gonz\'alez and P. S. Letelier, Phys.  Rev.  D {\bf 62}, 064025 (2000).

\bibitem{VL1} D. Vogt  and P. S. Letelier, Phys.  Rev.  D {\bf 68}, 084010 (2003).

\bibitem{UL1} M. Ujevic  and P. S. Letelier, Phys.  Rev.  D {\bf 70}, 084015 (2004).

\bibitem{LBZ} T. Ledvinka, J. Bi\u{c}\'{a}k, and M.  \u{Z}ofka, in {\it
Proceeding of 8th Marcel-Grossmann  Meeting in General Relativity}, edited by
T. Piran  (World  Scientific, Singapore, 1999) 

\bibitem{GG4} G. Garc\'\i a-Reyes and G. A. Gonz\'alez, 
Brazilian Journal of Physics {\bf 37}, no. 3B, 1094 (2007).

\bibitem{LET1} P. S. Letelier, Phys. Rev. D {\bf 60},  104042  (1999). 

\bibitem{KBL} J. Katz, J. Bi\u{c}\'ak, and D. Lynden-Bell, Class. Quantum Grav.
{\bf 16}, 4023 (1999).

\bibitem{GG1} G. Garc\'\i a R. and G. A. Gonz\'alez, Phys.  Rev. D  {\bf 69},
124002 (2004).

\bibitem{VL2} D. Vogt  and P. S. Letelier, Phys. Rev. D 70, 064003 (2004).

\bibitem{GG2} G. Garc\'\i a-Reyes and G. A. Gonz\'alez,  Class. Quantum Grav.
{\bf 21}, 4845 (2004).

\bibitem{GG3} G. Garc\'\i a-Reyes and G. A. Gonz\'alez, Phys.  Rev. D {\bf 70},
104005 (2004).

\bibitem{SYN} J. L. Synge, {\it Relativity: The General   Theory}.
(North-Holland, Amsterdam, 1966).

\bibitem{NM} G. Neugebauer and R. Meinel, Phys. Rev. Lett. {\bf 75}, 3046
(1995).

\bibitem{KLE1} C. Klein, Class. Quantum Grav. {\bf 14},  2267  (1997). 
\label{bib:KLE1}

\bibitem{KR} C. Klein and O. Richter, Phys. Rev. Lett.  {\bf 83}, 2884 (1999).

\bibitem{KLE2} C. Klein, Phys. Rev. D {\bf 63}, 064033  (2001).

\bibitem{FK} J. Frauendiener and C. Klein, Phys. Rev.  D {\bf 63}, 084025
(2001).

\bibitem{KLE3} C. Klein, Phys. Rev.  D {\bf 65}, 084029  (2002).

\bibitem{KLE4} C. Klein, Phys. Rev. D {\bf 68}, 027501  (2003).
\label{bib:KLE4}

\bibitem{KLE5} C. Klein, Ann. Phys. (Leipzig) {\bf 12}, 599  (2003).
\label{bib:KLE5}

\bibitem{KSHM} D. Kramer, H. Stephani, E. Herlt, and  M.   McCallum, {\it Exact
Solutions of Einsteins's  Field   Equations} (Cambridge University Press,
Cambridge,  England,  1980).


\bibitem{RGK} V. C. Rubin, J. A. Graham and J. D. P Kenney. Ap. J. {\bf 394}, L9, (1992).

\bibitem{RFF} H. Rix, M. Franx, D. Fisher and G. Illingworth. Ap. J. {\bf 400}, L5, (1992).

\bibitem{BER} F. Bertola {\it et al}. Ap. J. {\bf 458}, L67 (1996).

\bibitem{STRUCK} C. Struck, Phys. Rep. {\bf 321}, 1 (1999).

\bibitem{CBG} R. Ciri, D. Bettoni, and G. Galletta, 
Nature {\bf 375}, 661 (1995).

\bibitem{E2} F.J. Ernst, Phys. Rev. D {\bf 168}, 1415  (1968).


\bibitem{PH} A. Papapetrou and A. Hamouni, Ann. Inst. Henri Poincar\'e {\bf 9}, 179 (1968)

\bibitem{LICH} A. Lichnerowicz, C.R. Acad. Sci. {\bf 273}, 528 (1971) 

\bibitem{TAUB} A. H. Taub, J. Math. Phys. {\bf 21}, 1423 (1980)

\bibitem{IS1} E. Israel, Nuovo Cimento {\bf 44B}, 1 (1966)

\bibitem{IS2} E. Israel, Nuovo Cimento {\bf 48B}, 463 (1967)

\bibitem{POI} E. Poisson, {\it A Relativist's Toolkit: The Mathematics of Black-Hole Mechanics}. (Cambridge University Press, 2004)

\bibitem{CHAN} S. Chandrasekar, {\it The Mathematical Theory of Black Holes}. (Oxford University Press, 1992).

\bibitem{FLU} L.D. Landau and E.M. Lifshitz, {\it Fluid 
Mechanics}(Addison-Wesley, Reading, MA, 1989).

\bibitem{FHS} F. H. Seguin, Astrophys. J. {\bf 197}, 745 (1975).

\end{thebibliography}
\end{document}